\documentclass[aps, prb, reprint, groundscriptaddress]{revtex4-2}

\usepackage{amsmath, newtxtext, newtxmath} 
\usepackage{graphicx} 
\usepackage{hyperref}
\hypersetup{
    colorlinks = true,        
    linkcolor = blue,          
    citecolor = blue,        
    filecolor = magenta,      
    urlcolor = blue           
}
\usepackage{color}
\newcommand{\textbl}{\textcolor{black}}

\usepackage{braket} 
\usepackage{dcolumn} 
\usepackage{bm} 
\usepackage{soul} 

\begin{document}

\title{Single shot x-ray diffractometry in SACLA with pulsed magnetic fields up to 16 T}

\author{Akihiko~Ikeda}
\email[]{ikeda@issp.u-tokyo.ac.jp}
\author{Yasuhiro~H.~Matsuda}
\email[]{ymatsuda@issp.u-tokyo.ac.jp}
\author{Xuguang~Zhou}
\author{Takeshi~Yajima}
\affiliation{Institute for Solid State Physics, University of Tokyo, Kashiwa, Chiba, Japan}
\author{Yuya~Kubota}
\author{Kensuke~Tono}
\author{Makina~Yabashi}
\affiliation{RIKEN SPring-8 Center, Kouto, Hyogo, Japan}
\affiliation{Japan Synchrotron Radiation Research Institute, Kouto, Hyogo, Japan}

\date{\today}

\begin{abstract}
\textbl{Single shot x-ray diffraction (XRD) experiments have been performed with a x-ray free electron laser (XFEL) under pulsed high magnetic fields up to 16 T generated with a nondestructive minicoil.
The antiferromagnetic insulator phase in a perovskite manganaite, Pr$_{0.6}$Ca$_{0.4}$MnO$_{3}$, is collapsed at a magnetic field of $\approx 8$ T with an emergence of the ferromagnetic metallic phase, which is observed via the accompanying lattice changes in a series of the single shot XRD.
The feasibility of the single shot XRD experiment under ultrahigh magnetic fields beyond 100 T is discussed, which is generated with a portable destructive pulse magnet.}
\end{abstract}

\maketitle

\section{Introduction}
As a new light source, free electron lasers (XFELs) are prominently characterized by the ultrahigh transverse coherence, \textbl{ultrashort pulses between hundreds to a few}  femtoseconds and a high-photon flux of $10^{11-12}$ photons/pulse that realizes single shot experiments \cite{LCLS, Seddon}.
The advent of XFEL technique has provided us with such new experimental techniques to explore new area of science as coherent diffraction from a single nano-cluster, diffraction before destruction from a living cell, femtosecond time resolution pump-probe experiments, and also experiments with single shot x-ray diffractometry (XRD) under shocked compression environment.
\textbl{We are proposing an x-ray experiment at an extremely high magnetic field of above 100 T where we make use of} the ultrashort temporal width and the single shot experiment of a XFEL, and a destructive pulse magnet for 100 T generation.

High field experiments have been successfully combined with synchrotron radiation (SR) based XRD and spectroscopies for condensed matter experiments with DC magnets up to 15 T \cite{KiryukhinPRB, KiryukhinPRL, KiryukhinPRL2, NarumiJPSJ, NarumiJPSJ2} and non-destructive pulse magnets up to 50 T \cite{YHMatsudaPhysicaB, NarumiJSR, YHMatsudaPRL, Sikora, Ruff, YHMatsudaPRB, Islam, Billette, YHMatsudaJPSJ}.
Recent advent of single shot techniques with XFEL, further made possible the observations of the weak superlattice reflections of charge density wave order appearing in a cuprate by suppressing the high-$T_{\rm{C}}$ superconductivity up to 32 T \cite{Nojiri1, Nojiri2}.
These high-field studies with SR and XFEL have been confined below 50 T.
To generate magnetic fields over 50 T is a challenge for a portable non-destructive pulse magnet.
Non-destructive magnets generating above 50 T  are available only in specific high magnetic field facilities over the world, where pulsed magnetic fields up to 100 T are available with sufficient safety measures \cite{Battesti}.

For x-ray experiments at well above 50 T, we propose a use of a single turn coil (STC), a destructive pulse magnet,  instead of non-destructive pulse magnets, because a STC can be portable and generates magnetic fields over 100 T \cite{Herlach1971, Oliver01}.
Generating high fields well beyond 100 T \textbl{inevitably} requires destructive pulse magnets where magnetic pressure \textbl{beyond 4 GPa} destroys the coil.
For condensed matter experiments, flux compression and single turn coil techniques have been implemented \cite{Herlach, Miura, DN1200}.
The temporal duration of the generated field by a single turn coil is only a few micronseconds, which is about three orders of magnitudes smaller than a pulse of a non-destructive magnet.
It is still long enough for \textbl{the single shot experiment with an XFEL pulse}.
Note also an exceptional single shot technique utilizing SR from storage rings \cite{note1}, \textbl{which is potentially usable with a single shot pulses of 100 T of a few micronseconds.}

For a 100 T experiment at an XFEL site, one needs to construct a portable 100 T generator.
\textbl{It is profitable} to test equipments such as nonmetallic vacuum tubes and cryostats that are conventionally used with the STC experiments.
Portable STCs had been implemented before \cite{Oliver01, Herlach1971}.
A portable STC to be build should be less than 200 kg.
We are now constructing a portable pulse power with 30 kV charging and 4.5 kJ energy with a mono-spark gap.
Conventionally, two STCs are in operation in Institute for Solid State Physics (ISSP), University of Tokyo, with 200 kJ at 50 kV with a total weight of 6 tons.
In ISSP, specially prepared nonmetallic cryostat and vacuum system have been implemented for destructive pulse experiments at low temperatures.
These equipments are proof against the explosions of the STC magnet.
It is most preferable to use these equipments in the 100 T experiments in SPring-8 Angstrom Compact free electron LAser (SACLA).
To check this possibility, we need to test if these equipments are compatible with the x-ray experiments in SACLA.
Thus, it is fruitful to test the feasibility of those equipments at SACLA with a minicoil mimicking the 100 T experiment with a portable STC. 

\textbl{As preceding studies for the singe shot XRD at 100 T in SACLA,  we have discovered that the lattices of materials are actually changing under 100 T fields of a few micronseconds pulse generated in destructive pulse magnets \cite{IkedaMG}.
The direct observations of the lattice changes are performed using a high-speed strain sensor utilizing fiber Bragg grating (FBG), which is devised for magnetostriction measurements at above 100 T  \cite{IkedaFBG2017}.
A candidate for XRD study at 100 T is solid oxygen for structural analysis.
A high field beyond 120 T induces a new thermodynamic phase called $\theta$ phase \cite{NomuraPRL2014}.
A cubic lattice geometry is anticipated in the $\theta$ phase in contrast to the monoclinic $\alpha$ phase at low field phase, which remains elusive until a XRD is performed at 120 T.
An magnetostriction measurement found a first order lattice change accompanying the magnetization jump, indicating a lattice change at $\alpha$-$\theta$ transition.
Another candidate is a perovskite cobaltite LaCoO$_{3}$ for investigating the structure and electronic states at 100 T.
A peculiar spin crossover is found to be induced by magnetic fields beyond 100 T \cite{IkedaLCO, IkedaArxiv}.
There are two high field phases whose origins are controversially argued to be an Bose-Einstein condensation of excitons, a excitonic insulator, or a crystallization of the spin-state degree of freedom.
This is to be determined by observing a superlattice reflection from the spin-state crystallization and also by observing the $d$ electron states by the x-ray emission spectroscopy.
Besides them, there are plenty of magnetic phase transitions at ultrahigh magnetic fields where the lattice changes plays major roles \cite{MiyataPRL, TerashimaJPSJ, YMatsudaNC}.}

In this paper, we report an experiment of single shot XRD at SACLA with a non-destructive room-temperature-bore magnet mimicking the single turn coil.
The purpose of this work is to see the feasibility of an application of the ready-made explosion-proof equipments for the STC to XRD experiments at SACLA.
We successfully detect the field induced phase transition of Pr$_{0.6}$Ca$_{0.4}$MnO$_{3}$ with a series of single shot XRD with SACLA  up to 16 T.
We comment on a design for a portable single turn coil system for SACLA.

\section{Experiment}
\begin{figure}
\includegraphics[width = \linewidth, clip]{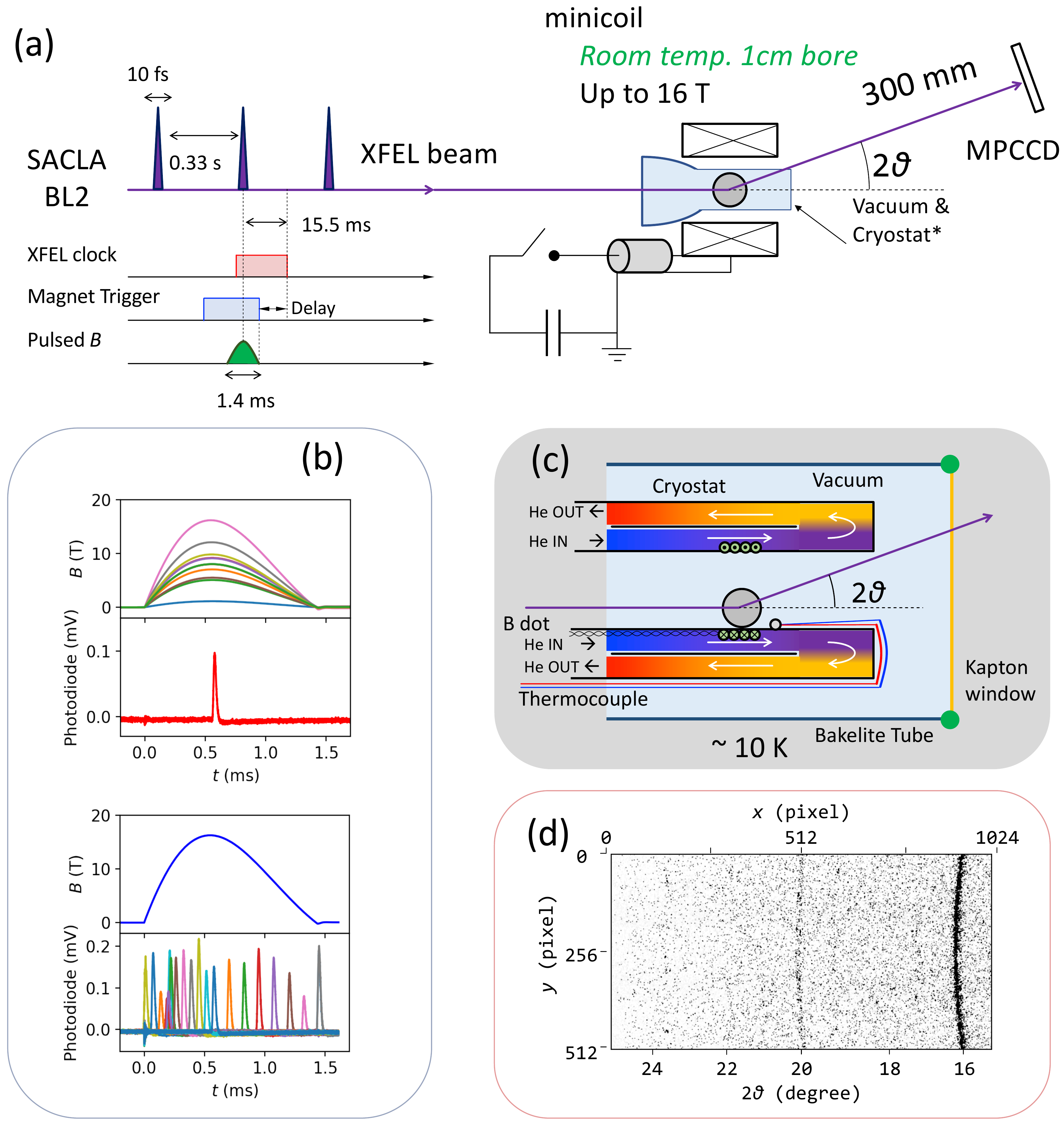} 
\caption{(a) A schematic drawing of the experimental setup of the single shot XRD in SACLA with a pulse magnet. The temporal diagram of the measurement is also shown. 
(b) The time evolution of the magnetic field and the XFEL pulse intensity. Magnetic fields can be varied in two ways in a single shot XRD.
The generated magnetic field is varied with a fixed XFEL timing on top of the pulsed magnetic field.
In another way, the XFEL timing is varied with a fixed waveform of the pulsed magnetic field. 
(c) A schematic drawing of a part of the nonmetallic cryostat used in ISSP for STC experiments, which is used this time for XRD study for verification of the compatibility.
(d) An image of a single shot XRD recorded by the MPCCD image sensor with a monochromatic beam of the XFEL.
\label{exp}}
\end{figure}

The experiment was performed at a hard x-ray beam line BL2, in SACLA  \cite{IshikawaNP, YabashiJSR}.
A schematic drawing of the experimental setup is shown in Fig. \ref{exp}(a).
The XFEL was tuned to 16 keV with a mean pulse energy of $\sim100$ $\mu$J/pulse.
Pink beam ($\Delta E/E\sim10^{-3}$) and monochromatic beam ($\Delta E/E\sim10^{-4}$) were used.
The single shot diffraction signals were monitored with a multi-port charge-coupled device (MPCCD) image sensor \cite{Kameshima}.
The magnetic fields were generated with a mini bank system of 2.4 kJ at 2000 V (1.2 mF) \cite{YHM2007, YHMatsudaPhysicaB} and a room temperature coil wounded by hand.
The waveforms of the pulsed magnetic fields are shown in Fig. \ref{exp}(b).

The bore of the coil is 1cm diameter and 1cm long in axial direction.
In the bore, a vacuum tube is suspended with a He-flow and non-metallic type cryostat located inside as shown in Fig. \ref{exp}(c).
The sample is put inside the cryostat, which is a position where a XFEL pulse hits and the diffracted signal escapes through Kapton windows at the back and top of the vacuum chamber.
The $2\theta$ range of $15 - 25$ degree is captured with a 25.6 $\times$ 51.2 mm window of the MPCCD with a distance of 300 mm from the sample to the detector.
The diffraction image is shown in Fig. \ref{exp} (d).

Pr$_{0.6}$Ca$_{0.4}$MnO$_{3}$ was powdered from a single crystal and dispersed in a glue whose effective thickness was $\sim10$ $\mu$m.
The powder was only roughly powdered on purpose so that spotty Debye ring is observed in XRD, where it is expected that we observe a speckle pattern around the diffraction spot reflecting the microstrain of a particle.

\begin{figure*}
\begin{center}
\includegraphics[width = \linewidth, clip]{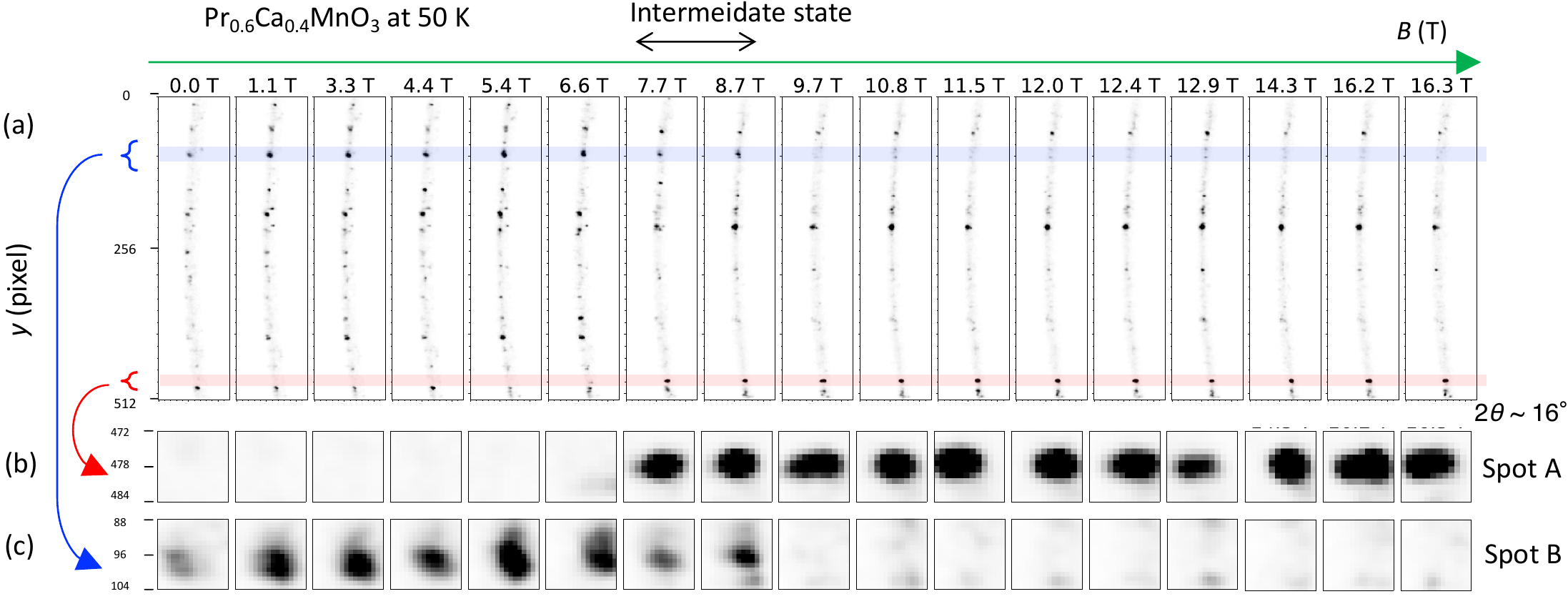} 
\caption{
(a) Image of a series of single shot XRD at $2\theta\simeq16^{\circ}$ with a pink beam at $h\nu = 16$ keV of XFEL in SACLA from a rough powder sample of Pr$_{0.6}$Ca$_{0.4}$MnO$_{3}$ at 50 K with various magnetic fields up to 16.3 T.
(b) Magnification of spot A and (c) spot B.
 \label{result}}
\end{center}
\end{figure*}

\begin{figure}[b]
\begin{center}
\includegraphics[width = 0.9\linewidth, clip]{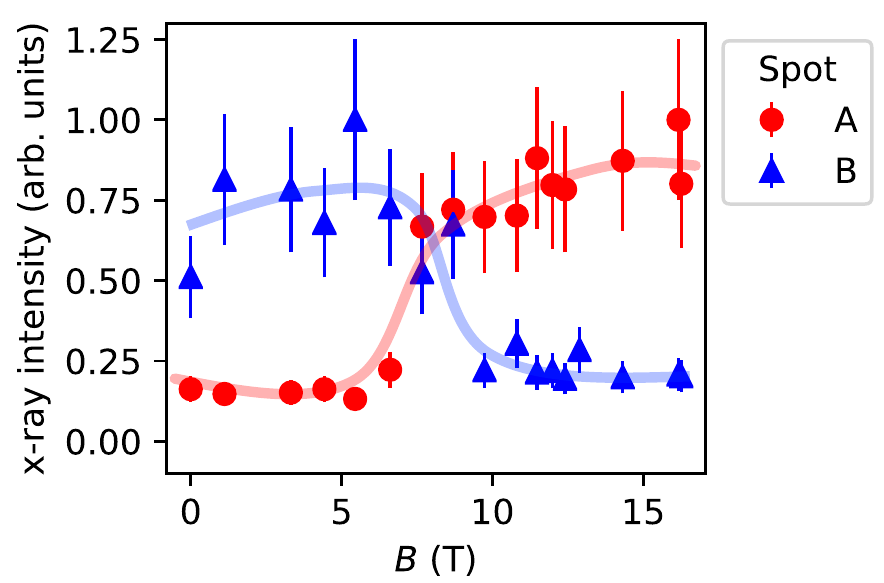} 
\caption{X-ray intensity of spot A and B of Figs. \ref{result}(b) and \ref{result}(c) as a function of external magnetic fields. \label{int}}
\end{center}
\end{figure}

\section{Result and discussion}
Fig. \ref{result} shows a series of single shot XRD at various magnetic fields up to 16.3 T at $2\theta \simeq16.3^{\circ}$, where 020, 200 and 112 reflections are relevant.
The spotty diffraction ring is produced from the roughly powdered sample.
One diffraction spot is coming from a single domain of a micro-particle of a sample.
The magnetic field effect is visible in Fig. \ref{result}.
\textbl{The trends of the images are similar to each other from 0 T to 6.6 T.}
The images from 9.7 T to 16.3 T have a common feature being distinct from the low field data.
The images at 7.7 T and 8.7 T seem to be linear combinations of both features from low field and high filed data, indicating a transient state.
Two magnifications are shown in Figs. \ref{result}(b) and \ref{result}(c).
In Fig. \ref{result}(b), it is clear that the spot A appears above 7.7 T.
In Fig. \ref{result}(c), on the contrary, the spot B disappears above 9.7 T.
The x-ray intensities of the spot A and spot B are shown as a function of $B$ as shown in Fig. \ref{int}.
The appearance of spot A means that the $\theta$-$2\theta$ configuration is satisfied after the field induced lattice parameter change of the particle for spot A.
In contrast, for spot B, which vanishes after the field induced lattice parameter change of the particle for spot B, the $\theta$-$2\theta$ configuration is no longer satisfied at high fields.
\textbl{Here we have used the pink beam, where the energy dispersion of 10$^{-3}$ results in a too large uncertainty in $2\theta$ for the assignment of the diffraction peaks.}

\begin{figure}[b]
\begin{center}
\includegraphics[width = 1\linewidth, clip]{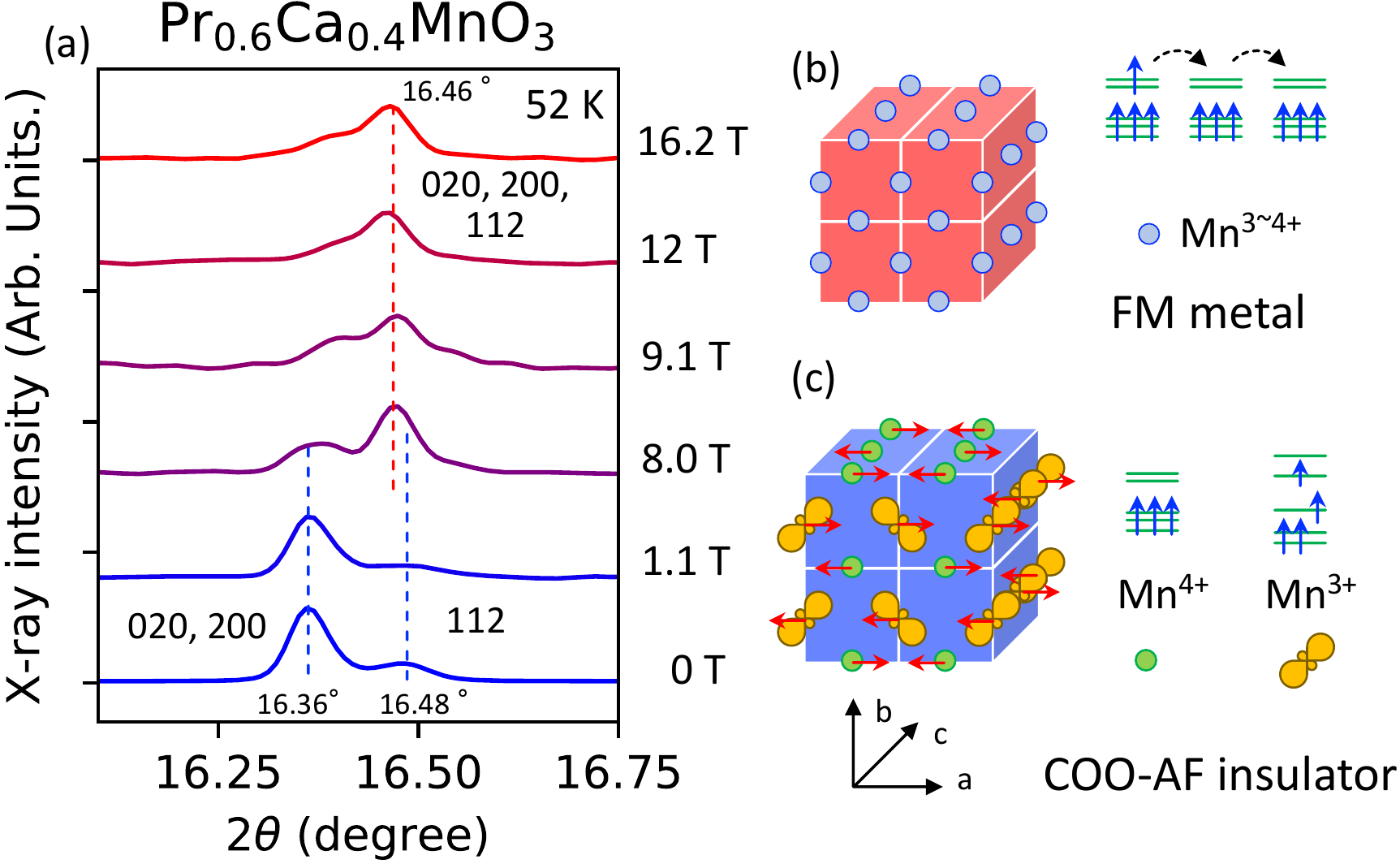} 
\caption{\textbl{(a) A series of single shot XRD at $2\theta\simeq16^{\circ}$ with a monochromated XFEL beam  at $h\nu = 16$ keV from a powdered sample of Pr$_{0.6}$Ca$_{0.4}$MnO$_{3}$ at 52 K at various magnetic fields up to 16.3 T. 
(b) A schematic drawing of the FM metallic phase of Pr$_{0.6}$Ca$_{0.4}$MnO$_{3}$ at above 8 T at 52 K.
It is depicted that the electron in the $e_{g}$ orbitals are delocalized and that the double exchange scheme favors the ferromagnetic spin configurations of the localized $t_{2g}$ orbitals, due to the Hund's coupling between the delocalized $e_{g}$ electrons and the localized $t_{2g}$ electrons.
(c) A schematic drawing of the COOI phase of Pr$_{0.6}$Ca$_{0.4}$MnO$_{3}$ at below 8 T at 52 K, depicting the charge, orbital and spin configurations, which is reproduced from Ref. \cite{Tokura}.
The localized electron occupies the $3z^{2} - r^{2}$ orbital at the Mn$^{3+}$ site, which is stable via the Jahn-Teller effect.
The antiferromagnetic spin configuration of so-called CE type is depicted.
 \label{mono}}
 }
\end{center}
\end{figure}

\textbl{For high resolution measurements in the diffraction angle, we have performed a series of single shot XRD at pulsed high magnetic fields with a monochromated XFEL beam.
The result is shown in Fig. \ref{mono}(a).
The diffraction peaks at 16.36$^{\circ}$ and 16.48$^{\circ}$ are identified to 020, 200 and, 112 reflections in a $Pbmn$ setting, respectively.
It is clear that the peak at 16.36$^{\circ}$ decreases at above 8 T, while the new peak at 16.46$^{\circ}$ increases.
This behavior is in good agreement with the previous result with SR \cite{YHMatsudaPhysicaB}.
In Pr$_{0.6}$Ca$_{0.4}$MnO$_{3}$, with decreasing temperature, the phase transition occurs from a ferromagnetic metallic (FM) phase to the charge and orbital ordered insulator (COOI) phase \cite{Tomioka}.
When COOI phase appears, the $a$ and $b$ axis elongates and the $c$ axis shrinks, stabilizing the occupied $3z^{2}-r^{2}$ orbital state \cite{Kuwahara, Tokura}.
The COOI phase is accompanied by antiferromagnetism.
Thereby, it can be collapsed with an external magnetic field.
With an emergence of the magnetic field induced FM phase, the $a$ and $b$ axis shrinks, and the $c$ axis elongates.
This is schematically depicted in Figs. \ref{mono}(b) and \ref{mono}(c).
According to this picture, the 020 and 200 diffractions are supposed to shift to higher $2\theta$ overlapping the peak of 112, and the 112 diffraction stays.
As a result, the diffraction peak at 16.38$^{\circ}$ decreases and the diffraction at 16.46$^{\circ}$ increases.
The observation in Fig. \ref{mono} is in good agreement with the above expectation, indicating that the field induced COOI with a single shot XRDs in SACLA is successfully observed.
}

\begin{figure}
\begin{center}
\includegraphics[width = 1.0\linewidth, clip]{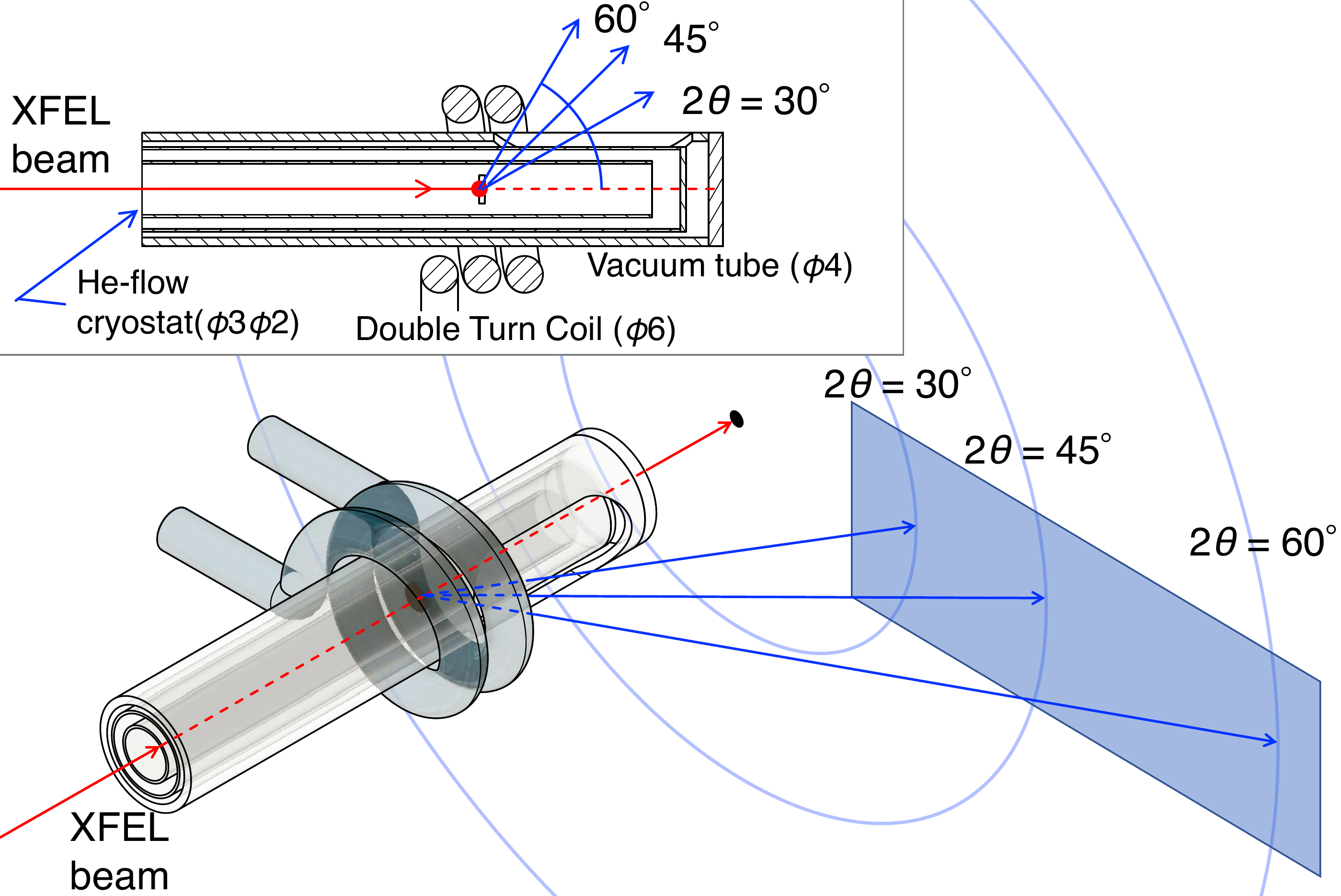} 
\caption{A schematic view of a proposed powder diffraction experiment in SACLA with a double turn coil, generating up to 80 T.
The diffraction angle $2\theta = 30$, 45, $60^{\circ}$ are shown.
The Debye ring is projected on a plane 1.4 cm away from the diffraction center.
The double turn coil, the vacuum tube and the He flow cryostat are made of copper, Kapton tube and FRP with a Kapton window for the diffracted x-ray.
The inset is a cross sectional view of the double turn coil, the vacuum tube and the cryostat, showing the opening angle for the diffracted x-ray.
\textbl{The shaded area colored in light blue represents the projection of the coverage by an IP with the size of $20 \times 40$ cm located 30 cm away from the diffraction center.}
 \label{sacla100}}
\end{center}
\end{figure}

So far it is shown that the single shot XRD is successfully conducted with a 1.4 ms-pulsed magnetic field and the non-metallic cryostat and the vacuum tube that are compatible with the STC experiments.
Also, the lattice change of Pr$_{0.6}$Ca$_{0.4}$MnO$_{3}$ accompanying the field induced phase transition is observed in a series  of single shot XRD experiments.
The observed changes of XRD are consistent with the shrinkage of $a$ and $b$ axis and an elongation of $c$ axis at high fields.

Here, we propose and discuss the feasibility of a single shot powder XRD measurement up to 100 T for a structural analysis.
In the present study, a rough powdered sample was employed with the motivation to observe  speckle patterns from a micrograin under stress in the two phase coexistence state during the phase transition.
However, such speckle patterns are not observed around the diffraction spots in Fig. \ref{result}.
The 100 T XRD experiment will focuses on the structural analysis by means of the smooth powder diffraction using a well powdered sample.
In the present study, a limited range of $2\theta < 25^{\circ}$ is covered, where a small number of low indexed diffraction peaks are obtained, being limited by the opening angles of the fiber-reinforced plastic (FRP) based cryostat and the minicoil.
Dimension of the present minicoil is $d=10$ mm in diameter and $h=10$ mm in axial length.
For structural analysis, it is favorable to cover a larger angle up to $2\theta \sim 60^{\circ}$.
A larger coverage of $2\theta$ becomes possible with a use of all-Kapton cryostat and a vacuum tube with a Kapton window as shown in Fig. \ref{sacla100}.
In the 100 T XRD experiment, a double turn coil is to be used which has a larger aspect ratio of $h/d = \sim 0.5$.
This allows us to cover $2\theta$ up to $60^{\circ}$ as schematically inspected in Fig. \ref{sacla100}. 

\textbl{Presently, the monochromated beam shows an appreciable resolution of XRD for the structural analysis, while the pink beam does not.
The use of the narrow and brighter beam of the seeded beam \cite{InoueNP} is even more practical in the 100 T XRD experiment.
Besides, use of a photon energy of 10 keV increases the photon number by a factor of $\sim6$ in stead of a photon energy of 16 keV used in the present study.
Presently, $h\nu = 16$ keV is used due to the limitation of the small $2\theta$ range.
With the larger $2\theta$ as proposed above, the use of $h\nu = 10$ keV is allowed.
A magnetic field of 80 T is estimated to be obtained with the double turn coil with a diameter of 6 mm and the portable bank system of ours currently under construction, which is rated at 30 kV with an energy of 4.5 kJ, generating 200 kA.
Instead of the MPCCD detector, we plan to use a conventional x-ray imaging plate (IP).
This is because an IP allows a much more robust detection against a fragmentation of the coil, which is not feasible with the fragile MPCCD detectors.}

The single crystalline experiment aiming a weak structure like superlattice reflection is at this moment not feasible due to the limited volume \textbl{($< 5 \times 5 \times 5$ mm$^{3}$)} of the magnetic field generated in STC, where sample rotation mechanism should be somehow installed.
Evaluation of this method will be a future task.




\section{Summary}
We performed single shot XRD experiments in SACLA with a pulsed magnetic filed up to 16 T.
We have successfully observed field induced lattice change in Pr$_{0.6}$Ca$_{0.4}$MnO$_{3}$, which is originated in a collapse of charge and orbital ordered insulator phase and appearance of a ferromagnetic metallic phase.
Based on the result, a methodology of a single shot XRD experiment up to 100 T range in SACLA using a portable 100 T generator is discussed.

\begin{acknowledgements}
This work was supported by JSPS KAKENHI Grant-in-Aid for Young Scientists  Grant No. 18K13493 and the Basic Science Program No. 18-001 of TEPCO memorial foundation. We thank Taka-hisa Arima for providing us with a high quality single crystalline sample.

\end{acknowledgements}

\bibliography{sacla}

\providecommand{\noopsort}[1]{}\providecommand{\singleletter}[1]{#1}%
\begin{thebibliography}{47}%
\makeatletter
\providecommand \@ifxundefined [1]{%
 \@ifx{#1\undefined}
}%
\providecommand \@ifnum [1]{%
 \ifnum #1\expandafter \@firstoftwo
 \else \expandafter \@secondoftwo
 \fi
}%
\providecommand \@ifx [1]{%
 \ifx #1\expandafter \@firstoftwo
 \else \expandafter \@secondoftwo
 \fi
}%
\providecommand \natexlab [1]{#1}%
\providecommand \enquote  [1]{``#1''}%
\providecommand \bibnamefont  [1]{#1}%
\providecommand \bibfnamefont [1]{#1}%
\providecommand \citenamefont [1]{#1}%
\providecommand \href@noop [0]{\@secondoftwo}%
\providecommand \href [0]{\begingroup \@sanitize@url \@href}%
\providecommand \@href[1]{\@@startlink{#1}\@@href}%
\providecommand \@@href[1]{\endgroup#1\@@endlink}%
\providecommand \@sanitize@url [0]{\catcode `\\12\catcode `\$12\catcode
  `\&12\catcode `\#12\catcode `\^12\catcode `\_12\catcode `\%12\relax}%
\providecommand \@@startlink[1]{}%
\providecommand \@@endlink[0]{}%
\providecommand \url  [0]{\begingroup\@sanitize@url \@url }%
\providecommand \@url [1]{\endgroup\@href {#1}{\urlprefix }}%
\providecommand \urlprefix  [0]{URL }%
\providecommand \Eprint [0]{\href }%
\providecommand \doibase [0]{https://doi.org/}%
\providecommand \selectlanguage [0]{\@gobble}%
\providecommand \bibinfo  [0]{\@secondoftwo}%
\providecommand \bibfield  [0]{\@secondoftwo}%
\providecommand \translation [1]{[#1]}%
\providecommand \BibitemOpen [0]{}%
\providecommand \bibitemStop [0]{}%
\providecommand \bibitemNoStop [0]{.\EOS\space}%
\providecommand \EOS [0]{\spacefactor3000\relax}%
\providecommand \BibitemShut  [1]{\csname bibitem#1\endcsname}%
\let\auto@bib@innerbib\@empty
\bibitem [{\citenamefont {Bostedt}\ \emph {et~al.}(2016)\citenamefont
  {Bostedt}, \citenamefont {Boutet}, \citenamefont {Fritz}, \citenamefont
  {Huang}, \citenamefont {Lee}, \citenamefont {Lemke}, \citenamefont {Robert},
  \citenamefont {Schlotter}, \citenamefont {Turner},\ and\ \citenamefont
  {Williams}}]{LCLS}%
  \BibitemOpen
  \bibfield  {author} {\bibinfo {author} {\bibfnamefont {C.}~\bibnamefont
  {Bostedt}}, \bibinfo {author} {\bibfnamefont {S.}~\bibnamefont {Boutet}},
  \bibinfo {author} {\bibfnamefont {D.~M.}\ \bibnamefont {Fritz}}, \bibinfo
  {author} {\bibfnamefont {Z.}~\bibnamefont {Huang}}, \bibinfo {author}
  {\bibfnamefont {H.~J.}\ \bibnamefont {Lee}}, \bibinfo {author} {\bibfnamefont
  {H.~T.}\ \bibnamefont {Lemke}}, \bibinfo {author} {\bibfnamefont
  {A.}~\bibnamefont {Robert}}, \bibinfo {author} {\bibfnamefont {W.~F.}\
  \bibnamefont {Schlotter}}, \bibinfo {author} {\bibfnamefont {J.~J.}\
  \bibnamefont {Turner}},\ and\ \bibinfo {author} {\bibfnamefont {G.~J.}\
  \bibnamefont {Williams}},\ }\href
  {https://doi.org/10.1103/RevModPhys.88.015007} {\bibfield  {journal}
  {\bibinfo  {journal} {Rev. Mod. Phys.}\ }\textbf {\bibinfo {volume} {88}},\
  \bibinfo {pages} {015007} (\bibinfo {year} {2016})}\BibitemShut {NoStop}%
\bibitem [{\citenamefont {Seddon}\ \emph {et~al.}(2017)\citenamefont {Seddon},
  \citenamefont {Clarke}, \citenamefont {Dunning}, \citenamefont
  {Masciovecchio}, \citenamefont {Milne}, \citenamefont {Parmigiani},
  \citenamefont {Rugg}, \citenamefont {Spence}, \citenamefont {Thompson},
  \citenamefont {Ueda}, \citenamefont {Vinko}, \citenamefont {Wark},\ and\
  \citenamefont {Wurth}}]{Seddon}%
  \BibitemOpen
  \bibfield  {author} {\bibinfo {author} {\bibfnamefont {E.~A.}\ \bibnamefont
  {Seddon}}, \bibinfo {author} {\bibfnamefont {J.~A.}\ \bibnamefont {Clarke}},
  \bibinfo {author} {\bibfnamefont {D.~J.}\ \bibnamefont {Dunning}}, \bibinfo
  {author} {\bibfnamefont {C.}~\bibnamefont {Masciovecchio}}, \bibinfo {author}
  {\bibfnamefont {C.~J.}\ \bibnamefont {Milne}}, \bibinfo {author}
  {\bibfnamefont {F.}~\bibnamefont {Parmigiani}}, \bibinfo {author}
  {\bibfnamefont {D.}~\bibnamefont {Rugg}}, \bibinfo {author} {\bibfnamefont
  {J.~C.~H.}\ \bibnamefont {Spence}}, \bibinfo {author} {\bibfnamefont {N.~R.}\
  \bibnamefont {Thompson}}, \bibinfo {author} {\bibfnamefont {K.}~\bibnamefont
  {Ueda}}, \bibinfo {author} {\bibfnamefont {S.~M.}\ \bibnamefont {Vinko}},
  \bibinfo {author} {\bibfnamefont {J.~S.}\ \bibnamefont {Wark}},\ and\
  \bibinfo {author} {\bibfnamefont {W.}~\bibnamefont {Wurth}},\ }\href
  {https://doi.org/10.1088/1361-6633/aa7cca} {\bibfield  {journal} {\bibinfo
  {journal} {Rep. Prog. Phys.}\ }\textbf {\bibinfo {volume} {80}},\ \bibinfo
  {pages} {115901} (\bibinfo {year} {2017})}\BibitemShut {NoStop}%
\bibitem [{\citenamefont {Kiryukhin}\ and\ \citenamefont
  {Keimer}(1995)}]{KiryukhinPRB}%
  \BibitemOpen
  \bibfield  {author} {\bibinfo {author} {\bibfnamefont {V.}~\bibnamefont
  {Kiryukhin}}\ and\ \bibinfo {author} {\bibfnamefont {B.}~\bibnamefont
  {Keimer}},\ }\href {https://doi.org/10.1103/PhysRevB.52.R704} {\bibfield
  {journal} {\bibinfo  {journal} {Phys. Rev. B}\ }\textbf {\bibinfo {volume}
  {52}},\ \bibinfo {pages} {R704} (\bibinfo {year} {1995})}\BibitemShut
  {NoStop}%
\bibitem [{\citenamefont {Kiryukhin}\ \emph {et~al.}(1995)\citenamefont
  {Kiryukhin}, \citenamefont {Keimer},\ and\ \citenamefont
  {Moncton}}]{KiryukhinPRL}%
  \BibitemOpen
  \bibfield  {author} {\bibinfo {author} {\bibfnamefont {V.}~\bibnamefont
  {Kiryukhin}}, \bibinfo {author} {\bibfnamefont {B.}~\bibnamefont {Keimer}},\
  and\ \bibinfo {author} {\bibfnamefont {D.~E.}\ \bibnamefont {Moncton}},\
  }\href {https://doi.org/10.1103/PhysRevLett.74.1669} {\bibfield  {journal}
  {\bibinfo  {journal} {Phys. Rev. Lett.}\ }\textbf {\bibinfo {volume} {74}},\
  \bibinfo {pages} {1669} (\bibinfo {year} {1995})}\BibitemShut {NoStop}%
\bibitem [{\citenamefont {Kiryukhin}\ \emph {et~al.}(1996)\citenamefont
  {Kiryukhin}, \citenamefont {Keimer}, \citenamefont {Hill},\ and\
  \citenamefont {Vigliante}}]{KiryukhinPRL2}%
  \BibitemOpen
  \bibfield  {author} {\bibinfo {author} {\bibfnamefont {V.}~\bibnamefont
  {Kiryukhin}}, \bibinfo {author} {\bibfnamefont {B.}~\bibnamefont {Keimer}},
  \bibinfo {author} {\bibfnamefont {J.~P.}\ \bibnamefont {Hill}},\ and\
  \bibinfo {author} {\bibfnamefont {A.}~\bibnamefont {Vigliante}},\ }\href
  {https://doi.org/10.1103/PhysRevLett.76.4608} {\bibfield  {journal} {\bibinfo
   {journal} {Phys. Rev. Lett.}\ }\textbf {\bibinfo {volume} {76}},\ \bibinfo
  {pages} {4608} (\bibinfo {year} {1996})}\BibitemShut {NoStop}%
\bibitem [{\citenamefont {Narumi}\ \emph {et~al.}(2004)\citenamefont {Narumi},
  \citenamefont {Katsumata}, \citenamefont {Tabata}, \citenamefont {Kimura},
  \citenamefont {Tanaka}, \citenamefont {Nakamura}, \citenamefont {Shimomura},
  \citenamefont {Matsuda}, \citenamefont {Harada}, \citenamefont {Nishiyama},
  \citenamefont {Ishikawa}, \citenamefont {Kitamura}, \citenamefont {Hara},
  \citenamefont {Tanaka}, \citenamefont {Tamasaku}, \citenamefont {Yabashi},
  \citenamefont {Goto}, \citenamefont {Ohashi}, \citenamefont {Takeshita},
  \citenamefont {Ohata}, \citenamefont {Matsushita},\ and\ \citenamefont
  {Bizen}}]{NarumiJPSJ}%
  \BibitemOpen
  \bibfield  {author} {\bibinfo {author} {\bibfnamefont {Y.}~\bibnamefont
  {Narumi}}, \bibinfo {author} {\bibfnamefont {K.}~\bibnamefont {Katsumata}},
  \bibinfo {author} {\bibfnamefont {Y.}~\bibnamefont {Tabata}}, \bibinfo
  {author} {\bibfnamefont {S.}~\bibnamefont {Kimura}}, \bibinfo {author}
  {\bibfnamefont {Y.}~\bibnamefont {Tanaka}}, \bibinfo {author} {\bibfnamefont
  {T.}~\bibnamefont {Nakamura}}, \bibinfo {author} {\bibfnamefont
  {S.}~\bibnamefont {Shimomura}}, \bibinfo {author} {\bibfnamefont
  {M.}~\bibnamefont {Matsuda}}, \bibinfo {author} {\bibfnamefont
  {I.}~\bibnamefont {Harada}}, \bibinfo {author} {\bibfnamefont
  {Y.}~\bibnamefont {Nishiyama}}, \bibinfo {author} {\bibfnamefont
  {T.}~\bibnamefont {Ishikawa}}, \bibinfo {author} {\bibfnamefont
  {H.}~\bibnamefont {Kitamura}}, \bibinfo {author} {\bibfnamefont
  {T.}~\bibnamefont {Hara}}, \bibinfo {author} {\bibfnamefont {T.}~\bibnamefont
  {Tanaka}}, \bibinfo {author} {\bibfnamefont {K.}~\bibnamefont {Tamasaku}},
  \bibinfo {author} {\bibfnamefont {M.}~\bibnamefont {Yabashi}}, \bibinfo
  {author} {\bibfnamefont {S.}~\bibnamefont {Goto}}, \bibinfo {author}
  {\bibfnamefont {H.}~\bibnamefont {Ohashi}}, \bibinfo {author} {\bibfnamefont
  {K.}~\bibnamefont {Takeshita}}, \bibinfo {author} {\bibfnamefont
  {T.}~\bibnamefont {Ohata}}, \bibinfo {author} {\bibfnamefont
  {T.}~\bibnamefont {Matsushita}},\ and\ \bibinfo {author} {\bibfnamefont
  {T.}~\bibnamefont {Bizen}},\ }\href {https://doi.org/10.1143/jpsj.73.2650}
  {\bibfield  {journal} {\bibinfo  {journal} {J. Phys. Soc. Jpn.}\ }\textbf
  {\bibinfo {volume} {73}},\ \bibinfo {pages} {2650} (\bibinfo {year}
  {2004})}\BibitemShut {NoStop}%
\bibitem [{\citenamefont {Narumi}\ \emph
  {et~al.}(2006{\natexlab{a}})\citenamefont {Narumi}, \citenamefont
  {Katsumata}, \citenamefont {Staub}, \citenamefont {Kindo}, \citenamefont
  {Kawauchi}, \citenamefont {Broennimann}, \citenamefont {Toyokawa},
  \citenamefont {Tanaka}, \citenamefont {Kikkawa}, \citenamefont {Yamamoto},
  \citenamefont {Hagiwara}, \citenamefont {Ishikawa},\ and\ \citenamefont
  {Kitamura}}]{NarumiJPSJ2}%
  \BibitemOpen
  \bibfield  {author} {\bibinfo {author} {\bibfnamefont {Y.}~\bibnamefont
  {Narumi}}, \bibinfo {author} {\bibfnamefont {K.}~\bibnamefont {Katsumata}},
  \bibinfo {author} {\bibfnamefont {U.}~\bibnamefont {Staub}}, \bibinfo
  {author} {\bibfnamefont {K.}~\bibnamefont {Kindo}}, \bibinfo {author}
  {\bibfnamefont {M.}~\bibnamefont {Kawauchi}}, \bibinfo {author}
  {\bibfnamefont {C.}~\bibnamefont {Broennimann}}, \bibinfo {author}
  {\bibfnamefont {H.}~\bibnamefont {Toyokawa}}, \bibinfo {author}
  {\bibfnamefont {Y.}~\bibnamefont {Tanaka}}, \bibinfo {author} {\bibfnamefont
  {A.}~\bibnamefont {Kikkawa}}, \bibinfo {author} {\bibfnamefont
  {T.}~\bibnamefont {Yamamoto}}, \bibinfo {author} {\bibfnamefont
  {M.}~\bibnamefont {Hagiwara}}, \bibinfo {author} {\bibfnamefont
  {T.}~\bibnamefont {Ishikawa}},\ and\ \bibinfo {author} {\bibfnamefont
  {H.}~\bibnamefont {Kitamura}},\ }\href
  {https://doi.org/10.1143/JPSJ.75.075001} {\bibfield  {journal} {\bibinfo
  {journal} {J. Phys. Soc. Jpn.}\ }\textbf {\bibinfo {volume} {75}},\ \bibinfo
  {pages} {075001} (\bibinfo {year} {2006}{\natexlab{a}})}\BibitemShut
  {NoStop}%
\bibitem [{\citenamefont {Matsuda}\ \emph {et~al.}(2004)\citenamefont
  {Matsuda}, \citenamefont {Ueda}, \citenamefont {Nojiri}, \citenamefont
  {Takahashi}, \citenamefont {Inami}, \citenamefont {Ohwada}, \citenamefont
  {Murakami},\ and\ \citenamefont {Arima}}]{YHMatsudaPhysicaB}%
  \BibitemOpen
  \bibfield  {author} {\bibinfo {author} {\bibfnamefont {Y.~H.}\ \bibnamefont
  {Matsuda}}, \bibinfo {author} {\bibfnamefont {Y.}~\bibnamefont {Ueda}},
  \bibinfo {author} {\bibfnamefont {H.}~\bibnamefont {Nojiri}}, \bibinfo
  {author} {\bibfnamefont {T.}~\bibnamefont {Takahashi}}, \bibinfo {author}
  {\bibfnamefont {T.}~\bibnamefont {Inami}}, \bibinfo {author} {\bibfnamefont
  {K.}~\bibnamefont {Ohwada}}, \bibinfo {author} {\bibfnamefont
  {Y.}~\bibnamefont {Murakami}},\ and\ \bibinfo {author} {\bibfnamefont
  {T.}~\bibnamefont {Arima}},\ }\href
  {https://doi.org/10.1016/j.physb.2004.01.139} {\bibfield  {journal} {\bibinfo
   {journal} {Physica B}\ }\textbf {\bibinfo {volume} {346}},\ \bibinfo {pages}
  {519} (\bibinfo {year} {2004})}\BibitemShut {NoStop}%
\bibitem [{\citenamefont {Narumi}\ \emph
  {et~al.}(2006{\natexlab{b}})\citenamefont {Narumi}, \citenamefont {Kindo},
  \citenamefont {Katsumata}, \citenamefont {Kawauchi}, \citenamefont
  {Broennimann}, \citenamefont {Staub}, \citenamefont {Toyokawa}, \citenamefont
  {Tanaka}, \citenamefont {Kikkawa}, \citenamefont {Yamamoto}, \citenamefont
  {Hagiwara}, \citenamefont {Ishikawa},\ and\ \citenamefont
  {Kitamura}}]{NarumiJSR}%
  \BibitemOpen
  \bibfield  {author} {\bibinfo {author} {\bibfnamefont {Y.}~\bibnamefont
  {Narumi}}, \bibinfo {author} {\bibfnamefont {K.}~\bibnamefont {Kindo}},
  \bibinfo {author} {\bibfnamefont {K.}~\bibnamefont {Katsumata}}, \bibinfo
  {author} {\bibfnamefont {M.}~\bibnamefont {Kawauchi}}, \bibinfo {author}
  {\bibfnamefont {C.}~\bibnamefont {Broennimann}}, \bibinfo {author}
  {\bibfnamefont {U.}~\bibnamefont {Staub}}, \bibinfo {author} {\bibfnamefont
  {H.}~\bibnamefont {Toyokawa}}, \bibinfo {author} {\bibfnamefont
  {Y.}~\bibnamefont {Tanaka}}, \bibinfo {author} {\bibfnamefont
  {A.}~\bibnamefont {Kikkawa}}, \bibinfo {author} {\bibfnamefont
  {T.}~\bibnamefont {Yamamoto}}, \bibinfo {author} {\bibfnamefont
  {M.}~\bibnamefont {Hagiwara}}, \bibinfo {author} {\bibfnamefont
  {T.}~\bibnamefont {Ishikawa}},\ and\ \bibinfo {author} {\bibfnamefont
  {H.}~\bibnamefont {Kitamura}},\ }\href
  {https://doi.org/10.1107/S0909049506006972} {\bibfield  {journal} {\bibinfo
  {journal} {J. Synchrotron Rad.}\ }\textbf {\bibinfo {volume} {13}},\ \bibinfo
  {pages} {271} (\bibinfo {year} {2006}{\natexlab{b}})}\BibitemShut {NoStop}%
\bibitem [{\citenamefont {Matsuda}\ \emph {et~al.}(2009)\citenamefont
  {Matsuda}, \citenamefont {Ouyang}, \citenamefont {Nojiri}, \citenamefont
  {Inami}, \citenamefont {Ohwada}, \citenamefont {Suzuki}, \citenamefont
  {Kawamura}, \citenamefont {Mitsuda},\ and\ \citenamefont
  {Wada}}]{YHMatsudaPRL}%
  \BibitemOpen
  \bibfield  {author} {\bibinfo {author} {\bibfnamefont {Y.~H.}\ \bibnamefont
  {Matsuda}}, \bibinfo {author} {\bibfnamefont {Z.~W.}\ \bibnamefont {Ouyang}},
  \bibinfo {author} {\bibfnamefont {H.}~\bibnamefont {Nojiri}}, \bibinfo
  {author} {\bibfnamefont {T.}~\bibnamefont {Inami}}, \bibinfo {author}
  {\bibfnamefont {K.}~\bibnamefont {Ohwada}}, \bibinfo {author} {\bibfnamefont
  {M.}~\bibnamefont {Suzuki}}, \bibinfo {author} {\bibfnamefont
  {N.}~\bibnamefont {Kawamura}}, \bibinfo {author} {\bibfnamefont
  {A.}~\bibnamefont {Mitsuda}},\ and\ \bibinfo {author} {\bibfnamefont
  {H.}~\bibnamefont {Wada}},\ }\href
  {https://doi.org/10.1103/PhysRevLett.103.046402} {\bibfield  {journal}
  {\bibinfo  {journal} {Phys. Rev. Lett.}\ }\textbf {\bibinfo {volume} {103}},\
  \bibinfo {pages} {046402} (\bibinfo {year} {2009})}\BibitemShut {NoStop}%
\bibitem [{\citenamefont {Sikora}\ \emph {et~al.}(2009)\citenamefont {Sikora},
  \citenamefont {Mathon}, \citenamefont {van~der Linden}, \citenamefont
  {Michalik}, \citenamefont {de~Teresa}, \citenamefont {Kapusta},\ and\
  \citenamefont {Pascarelli}}]{Sikora}%
  \BibitemOpen
  \bibfield  {author} {\bibinfo {author} {\bibfnamefont {M.}~\bibnamefont
  {Sikora}}, \bibinfo {author} {\bibfnamefont {O.}~\bibnamefont {Mathon}},
  \bibinfo {author} {\bibfnamefont {P.}~\bibnamefont {van~der Linden}},
  \bibinfo {author} {\bibfnamefont {J.~M.}\ \bibnamefont {Michalik}}, \bibinfo
  {author} {\bibfnamefont {J.~M.}\ \bibnamefont {de~Teresa}}, \bibinfo {author}
  {\bibfnamefont {C.}~\bibnamefont {Kapusta}},\ and\ \bibinfo {author}
  {\bibfnamefont {S.}~\bibnamefont {Pascarelli}},\ }\href
  {https://doi.org/10.1103/PhysRevB.79.220402} {\bibfield  {journal} {\bibinfo
  {journal} {Phys. Rev. B}\ }\textbf {\bibinfo {volume} {79}},\ \bibinfo
  {pages} {220402} (\bibinfo {year} {2009})}\BibitemShut {NoStop}%
\bibitem [{\citenamefont {Ruff}\ \emph {et~al.}(2010)\citenamefont {Ruff},
  \citenamefont {Islam}, \citenamefont {Clancy}, \citenamefont {Ross},
  \citenamefont {Nojiri}, \citenamefont {Matsuda}, \citenamefont {Dabkowska},
  \citenamefont {Dabkowski},\ and\ \citenamefont {Gaulin}}]{Ruff}%
  \BibitemOpen
  \bibfield  {author} {\bibinfo {author} {\bibfnamefont {J.~P.}\ \bibnamefont
  {Ruff}}, \bibinfo {author} {\bibfnamefont {Z.}~\bibnamefont {Islam}},
  \bibinfo {author} {\bibfnamefont {J.~P.}\ \bibnamefont {Clancy}}, \bibinfo
  {author} {\bibfnamefont {K.~A.}\ \bibnamefont {Ross}}, \bibinfo {author}
  {\bibfnamefont {H.}~\bibnamefont {Nojiri}}, \bibinfo {author} {\bibfnamefont
  {Y.~H.}\ \bibnamefont {Matsuda}}, \bibinfo {author} {\bibfnamefont {H.~A.}\
  \bibnamefont {Dabkowska}}, \bibinfo {author} {\bibfnamefont {A.~D.}\
  \bibnamefont {Dabkowski}},\ and\ \bibinfo {author} {\bibfnamefont {B.~D.}\
  \bibnamefont {Gaulin}},\ }\href
  {https://doi.org/10.1103/PhysRevLett.105.077203} {\bibfield  {journal}
  {\bibinfo  {journal} {Phys. Rev. Lett.}\ }\textbf {\bibinfo {volume} {105}},\
  \bibinfo {pages} {077203} (\bibinfo {year} {2010})}\BibitemShut {NoStop}%
\bibitem [{\citenamefont {Matsuda}\ \emph {et~al.}(2011)\citenamefont
  {Matsuda}, \citenamefont {Her}, \citenamefont {Michimura}, \citenamefont
  {Inami}, \citenamefont {Suzuki}, \citenamefont {Kawamura}, \citenamefont
  {Mizumaki}, \citenamefont {Kindo}, \citenamefont {Yamauara},\ and\
  \citenamefont {Hiroi}}]{YHMatsudaPRB}%
  \BibitemOpen
  \bibfield  {author} {\bibinfo {author} {\bibfnamefont {Y.~H.}\ \bibnamefont
  {Matsuda}}, \bibinfo {author} {\bibfnamefont {J.~L.}\ \bibnamefont {Her}},
  \bibinfo {author} {\bibfnamefont {S.}~\bibnamefont {Michimura}}, \bibinfo
  {author} {\bibfnamefont {T.}~\bibnamefont {Inami}}, \bibinfo {author}
  {\bibfnamefont {M.}~\bibnamefont {Suzuki}}, \bibinfo {author} {\bibfnamefont
  {N.}~\bibnamefont {Kawamura}}, \bibinfo {author} {\bibfnamefont
  {M.}~\bibnamefont {Mizumaki}}, \bibinfo {author} {\bibfnamefont
  {K.}~\bibnamefont {Kindo}}, \bibinfo {author} {\bibfnamefont
  {J.}~\bibnamefont {Yamauara}},\ and\ \bibinfo {author} {\bibfnamefont
  {Z.}~\bibnamefont {Hiroi}},\ }\href
  {https://doi.org/10.1103/PhysRevB.84.174431} {\bibfield  {journal} {\bibinfo
  {journal} {Phys. Rev. B}\ }\textbf {\bibinfo {volume} {84}},\ \bibinfo
  {pages} {174431} (\bibinfo {year} {2011})}\BibitemShut {NoStop}%
\bibitem [{\citenamefont {Islam}\ \emph {et~al.}(2012)\citenamefont {Islam},
  \citenamefont {Capatina}, \citenamefont {Ruff}, \citenamefont {Das},
  \citenamefont {Trakhtenberg}, \citenamefont {Nojiri}, \citenamefont {Narumi},
  \citenamefont {Welp},\ and\ \citenamefont {Canfield}}]{Islam}%
  \BibitemOpen
  \bibfield  {author} {\bibinfo {author} {\bibfnamefont {Z.}~\bibnamefont
  {Islam}}, \bibinfo {author} {\bibfnamefont {D.}~\bibnamefont {Capatina}},
  \bibinfo {author} {\bibfnamefont {J.~P.}\ \bibnamefont {Ruff}}, \bibinfo
  {author} {\bibfnamefont {R.~K.}\ \bibnamefont {Das}}, \bibinfo {author}
  {\bibfnamefont {E.}~\bibnamefont {Trakhtenberg}}, \bibinfo {author}
  {\bibfnamefont {H.}~\bibnamefont {Nojiri}}, \bibinfo {author} {\bibfnamefont
  {Y.}~\bibnamefont {Narumi}}, \bibinfo {author} {\bibfnamefont
  {U.}~\bibnamefont {Welp}},\ and\ \bibinfo {author} {\bibfnamefont {P.~C.}\
  \bibnamefont {Canfield}},\ }\href {https://doi.org/10.1063/1.3688251}
  {\bibfield  {journal} {\bibinfo  {journal} {Rev. Sci. Instrum.}\ }\textbf
  {\bibinfo {volume} {83}},\ \bibinfo {pages} {035101} (\bibinfo {year}
  {2012})}\BibitemShut {NoStop}%
\bibitem [{\citenamefont {Billette}\ \emph {et~al.}(2012)\citenamefont
  {Billette}, \citenamefont {Duc}, \citenamefont {Frings}, \citenamefont
  {Nardone}, \citenamefont {Zitouni}, \citenamefont {Detlefs}, \citenamefont
  {Roth}, \citenamefont {Crichton}, \citenamefont {Lorenzo},\ and\
  \citenamefont {Rikken}}]{Billette}%
  \BibitemOpen
  \bibfield  {author} {\bibinfo {author} {\bibfnamefont {J.}~\bibnamefont
  {Billette}}, \bibinfo {author} {\bibfnamefont {F.}~\bibnamefont {Duc}},
  \bibinfo {author} {\bibfnamefont {P.}~\bibnamefont {Frings}}, \bibinfo
  {author} {\bibfnamefont {M.}~\bibnamefont {Nardone}}, \bibinfo {author}
  {\bibfnamefont {A.}~\bibnamefont {Zitouni}}, \bibinfo {author} {\bibfnamefont
  {C.}~\bibnamefont {Detlefs}}, \bibinfo {author} {\bibfnamefont
  {T.}~\bibnamefont {Roth}}, \bibinfo {author} {\bibfnamefont {W.}~\bibnamefont
  {Crichton}}, \bibinfo {author} {\bibfnamefont {J.~E.}\ \bibnamefont
  {Lorenzo}},\ and\ \bibinfo {author} {\bibfnamefont {G.~L.}\ \bibnamefont
  {Rikken}},\ }\href {https://doi.org/10.1063/1.3701830} {\bibfield  {journal}
  {\bibinfo  {journal} {Rev. Sci. Instrum.}\ }\textbf {\bibinfo {volume}
  {83}},\ \bibinfo {pages} {043904} (\bibinfo {year} {2012})}\BibitemShut
  {NoStop}%
\bibitem [{\citenamefont {H.~Matsuda}\ and\ \citenamefont
  {Inami}(2013)}]{YHMatsudaJPSJ}%
  \BibitemOpen
  \bibfield  {author} {\bibinfo {author} {\bibfnamefont {Y.}~\bibnamefont
  {H.~Matsuda}}\ and\ \bibinfo {author} {\bibfnamefont {T.}~\bibnamefont
  {Inami}},\ }\href {https://doi.org/10.7566/JPSJ.82.021009} {\bibfield
  {journal} {\bibinfo  {journal} {J. Phys. Soc. Jpn.}\ }\textbf {\bibinfo
  {volume} {82}},\ \bibinfo {pages} {021009} (\bibinfo {year}
  {2013})}\BibitemShut {NoStop}%
\bibitem [{\citenamefont {Gerber}\ \emph {et~al.}(2015)\citenamefont {Gerber},
  \citenamefont {Jang}, \citenamefont {Nojiri}, \citenamefont {Matsuzawa},
  \citenamefont {Yasumura}, \citenamefont {Bonn}, \citenamefont {Liang},
  \citenamefont {Hardy}, \citenamefont {Islam}, \citenamefont {Mehta},
  \citenamefont {Song}, \citenamefont {Sikorski}, \citenamefont {Stefanescu},
  \citenamefont {Feng}, \citenamefont {Kivelson}, \citenamefont {Devereaux},
  \citenamefont {Shen}, \citenamefont {Kao}, \citenamefont {Lee}, \citenamefont
  {Zhu},\ and\ \citenamefont {Lee}}]{Nojiri1}%
  \BibitemOpen
  \bibfield  {author} {\bibinfo {author} {\bibfnamefont {S.}~\bibnamefont
  {Gerber}}, \bibinfo {author} {\bibfnamefont {H.}~\bibnamefont {Jang}},
  \bibinfo {author} {\bibfnamefont {H.}~\bibnamefont {Nojiri}}, \bibinfo
  {author} {\bibfnamefont {S.}~\bibnamefont {Matsuzawa}}, \bibinfo {author}
  {\bibfnamefont {H.}~\bibnamefont {Yasumura}}, \bibinfo {author}
  {\bibfnamefont {D.~A.}\ \bibnamefont {Bonn}}, \bibinfo {author}
  {\bibfnamefont {R.}~\bibnamefont {Liang}}, \bibinfo {author} {\bibfnamefont
  {W.~N.}\ \bibnamefont {Hardy}}, \bibinfo {author} {\bibfnamefont
  {Z.}~\bibnamefont {Islam}}, \bibinfo {author} {\bibfnamefont
  {A.}~\bibnamefont {Mehta}}, \bibinfo {author} {\bibfnamefont
  {S.}~\bibnamefont {Song}}, \bibinfo {author} {\bibfnamefont {M.}~\bibnamefont
  {Sikorski}}, \bibinfo {author} {\bibfnamefont {D.}~\bibnamefont
  {Stefanescu}}, \bibinfo {author} {\bibfnamefont {Y.}~\bibnamefont {Feng}},
  \bibinfo {author} {\bibfnamefont {S.~A.}\ \bibnamefont {Kivelson}}, \bibinfo
  {author} {\bibfnamefont {T.~P.}\ \bibnamefont {Devereaux}}, \bibinfo {author}
  {\bibfnamefont {Z.~X.}\ \bibnamefont {Shen}}, \bibinfo {author}
  {\bibfnamefont {C.~C.}\ \bibnamefont {Kao}}, \bibinfo {author} {\bibfnamefont
  {W.~S.}\ \bibnamefont {Lee}}, \bibinfo {author} {\bibfnamefont
  {D.}~\bibnamefont {Zhu}},\ and\ \bibinfo {author} {\bibfnamefont {J.~S.}\
  \bibnamefont {Lee}},\ }\href {https://doi.org/10.1126/science.aac6257}
  {\bibfield  {journal} {\bibinfo  {journal} {Science}\ }\textbf {\bibinfo
  {volume} {350}},\ \bibinfo {pages} {949} (\bibinfo {year}
  {2015})}\BibitemShut {NoStop}%
\bibitem [{\citenamefont {Jang}\ \emph {et~al.}(2016)\citenamefont {Jang},
  \citenamefont {Lee}, \citenamefont {Nojiri}, \citenamefont {Matsuzawa},
  \citenamefont {Yasumura}, \citenamefont {Nie}, \citenamefont {Maharaj},
  \citenamefont {Gerber}, \citenamefont {Liu}, \citenamefont {Mehta},
  \citenamefont {Bonn}, \citenamefont {Liang}, \citenamefont {Hardy},
  \citenamefont {Burns}, \citenamefont {Islam}, \citenamefont {Song},
  \citenamefont {Hastings}, \citenamefont {Devereaux}, \citenamefont {Shen},
  \citenamefont {Kivelson}, \citenamefont {Kao}, \citenamefont {Zhu},\ and\
  \citenamefont {Lee}}]{Nojiri2}%
  \BibitemOpen
  \bibfield  {author} {\bibinfo {author} {\bibfnamefont {H.}~\bibnamefont
  {Jang}}, \bibinfo {author} {\bibfnamefont {W.~S.}\ \bibnamefont {Lee}},
  \bibinfo {author} {\bibfnamefont {H.}~\bibnamefont {Nojiri}}, \bibinfo
  {author} {\bibfnamefont {S.}~\bibnamefont {Matsuzawa}}, \bibinfo {author}
  {\bibfnamefont {H.}~\bibnamefont {Yasumura}}, \bibinfo {author}
  {\bibfnamefont {L.}~\bibnamefont {Nie}}, \bibinfo {author} {\bibfnamefont
  {A.~V.}\ \bibnamefont {Maharaj}}, \bibinfo {author} {\bibfnamefont
  {S.}~\bibnamefont {Gerber}}, \bibinfo {author} {\bibfnamefont {Y.~J.}\
  \bibnamefont {Liu}}, \bibinfo {author} {\bibfnamefont {A.}~\bibnamefont
  {Mehta}}, \bibinfo {author} {\bibfnamefont {D.~A.}\ \bibnamefont {Bonn}},
  \bibinfo {author} {\bibfnamefont {R.}~\bibnamefont {Liang}}, \bibinfo
  {author} {\bibfnamefont {W.~N.}\ \bibnamefont {Hardy}}, \bibinfo {author}
  {\bibfnamefont {C.~A.}\ \bibnamefont {Burns}}, \bibinfo {author}
  {\bibfnamefont {Z.}~\bibnamefont {Islam}}, \bibinfo {author} {\bibfnamefont
  {S.}~\bibnamefont {Song}}, \bibinfo {author} {\bibfnamefont {J.}~\bibnamefont
  {Hastings}}, \bibinfo {author} {\bibfnamefont {T.~P.}\ \bibnamefont
  {Devereaux}}, \bibinfo {author} {\bibfnamefont {Z.~X.}\ \bibnamefont {Shen}},
  \bibinfo {author} {\bibfnamefont {S.~A.}\ \bibnamefont {Kivelson}}, \bibinfo
  {author} {\bibfnamefont {C.~C.}\ \bibnamefont {Kao}}, \bibinfo {author}
  {\bibfnamefont {D.}~\bibnamefont {Zhu}},\ and\ \bibinfo {author}
  {\bibfnamefont {J.~S.}\ \bibnamefont {Lee}},\ }\href
  {https://doi.org/10.1073/pnas.1612849113} {\bibfield  {journal} {\bibinfo
  {journal} {Proc. Natl. Acad. Sci. USA}\ }\textbf {\bibinfo {volume} {113}},\
  \bibinfo {pages} {14645} (\bibinfo {year} {2016})}\BibitemShut {NoStop}%
\bibitem [{\citenamefont {Battesti}\ \emph {et~al.}(2018)\citenamefont
  {Battesti}, \citenamefont {Beard}, \citenamefont {B\"{o}ser}, \citenamefont
  {Bruyant}, \citenamefont {Budker}, \citenamefont {Crooker}, \citenamefont
  {Daw}, \citenamefont {Flambaum}, \citenamefont {Inada}, \citenamefont
  {Irastorza}, \citenamefont {Karbstein}, \citenamefont {Kim}, \citenamefont
  {Kozlov}, \citenamefont {Melhem}, \citenamefont {Phipps}, \citenamefont
  {Pugnat}, \citenamefont {Rikken}, \citenamefont {Rizzo}, \citenamefont
  {Schott}, \citenamefont {Semertzidis}, \citenamefont {ten Kate},\ and\
  \citenamefont {Zavattini}}]{Battesti}%
  \BibitemOpen
  \bibfield  {author} {\bibinfo {author} {\bibfnamefont {R.}~\bibnamefont
  {Battesti}}, \bibinfo {author} {\bibfnamefont {J.}~\bibnamefont {Beard}},
  \bibinfo {author} {\bibfnamefont {S.}~\bibnamefont {B\"{o}ser}}, \bibinfo
  {author} {\bibfnamefont {N.}~\bibnamefont {Bruyant}}, \bibinfo {author}
  {\bibfnamefont {D.}~\bibnamefont {Budker}}, \bibinfo {author} {\bibfnamefont
  {S.~A.}\ \bibnamefont {Crooker}}, \bibinfo {author} {\bibfnamefont {E.~J.}\
  \bibnamefont {Daw}}, \bibinfo {author} {\bibfnamefont {V.~V.}\ \bibnamefont
  {Flambaum}}, \bibinfo {author} {\bibfnamefont {T.}~\bibnamefont {Inada}},
  \bibinfo {author} {\bibfnamefont {I.~G.}\ \bibnamefont {Irastorza}}, \bibinfo
  {author} {\bibfnamefont {F.}~\bibnamefont {Karbstein}}, \bibinfo {author}
  {\bibfnamefont {D.~L.}\ \bibnamefont {Kim}}, \bibinfo {author} {\bibfnamefont
  {M.~G.}\ \bibnamefont {Kozlov}}, \bibinfo {author} {\bibfnamefont
  {Z.}~\bibnamefont {Melhem}}, \bibinfo {author} {\bibfnamefont
  {A.}~\bibnamefont {Phipps}}, \bibinfo {author} {\bibfnamefont
  {P.}~\bibnamefont {Pugnat}}, \bibinfo {author} {\bibfnamefont
  {G.}~\bibnamefont {Rikken}}, \bibinfo {author} {\bibfnamefont
  {C.}~\bibnamefont {Rizzo}}, \bibinfo {author} {\bibfnamefont
  {M.}~\bibnamefont {Schott}}, \bibinfo {author} {\bibfnamefont {Y.~K.}\
  \bibnamefont {Semertzidis}}, \bibinfo {author} {\bibfnamefont {H.~H.~J.}\
  \bibnamefont {ten Kate}},\ and\ \bibinfo {author} {\bibfnamefont
  {G.}~\bibnamefont {Zavattini}},\ }\href
  {https://doi.org/10.1016/j.physrep.2018.07.005} {\bibfield  {journal}
  {\bibinfo  {journal} {Phys. Rep.}\ }\textbf {\bibinfo {volume} {765-766}},\
  \bibinfo {pages} {1} (\bibinfo {year} {2018})}\BibitemShut {NoStop}%
\bibitem [{\citenamefont {Herlach}\ \emph {et~al.}(1971)\citenamefont
  {Herlach}, \citenamefont {McBroom}, \citenamefont {Erber}, \citenamefont
  {Murray},\ and\ \citenamefont {Gearhart}}]{Herlach1971}%
  \BibitemOpen
  \bibfield  {author} {\bibinfo {author} {\bibfnamefont {F.}~\bibnamefont
  {Herlach}}, \bibinfo {author} {\bibfnamefont {R.}~\bibnamefont {McBroom}},
  \bibinfo {author} {\bibfnamefont {T.}~\bibnamefont {Erber}}, \bibinfo
  {author} {\bibfnamefont {J.}~\bibnamefont {Murray}},\ and\ \bibinfo {author}
  {\bibfnamefont {R.}~\bibnamefont {Gearhart}},\ }\href
  {https://doi.org/10.1109/tns.1971.4326194} {\bibfield  {journal} {\bibinfo
  {journal} {IEEE Trans. Nucl. Sci.}\ }\textbf {\bibinfo {volume} {18}},\
  \bibinfo {pages} {809} (\bibinfo {year} {1971})}\BibitemShut {NoStop}%
\bibitem [{\citenamefont {Portugall}\ \emph {et~al.}(1997)\citenamefont
  {Portugall}, \citenamefont {Puhlmann}, \citenamefont {M\"{u}ller},
  \citenamefont {Barczewski}, \citenamefont {Stolpe}, \citenamefont {Thiede},
  \citenamefont {Scholz}, \citenamefont {Ortenberg},\ and\ \citenamefont
  {Herlach}}]{Oliver01}%
  \BibitemOpen
  \bibfield  {author} {\bibinfo {author} {\bibfnamefont {O.}~\bibnamefont
  {Portugall}}, \bibinfo {author} {\bibfnamefont {N.}~\bibnamefont {Puhlmann}},
  \bibinfo {author} {\bibfnamefont {H.~U.}\ \bibnamefont {M\"{u}ller}},
  \bibinfo {author} {\bibfnamefont {M.}~\bibnamefont {Barczewski}}, \bibinfo
  {author} {\bibfnamefont {I.}~\bibnamefont {Stolpe}}, \bibinfo {author}
  {\bibfnamefont {M.}~\bibnamefont {Thiede}}, \bibinfo {author} {\bibfnamefont
  {H.}~\bibnamefont {Scholz}}, \bibinfo {author} {\bibfnamefont {M.~v.}\
  \bibnamefont {Ortenberg}},\ and\ \bibinfo {author} {\bibfnamefont
  {F.}~\bibnamefont {Herlach}},\ }\href
  {https://doi.org/10.1088/0022-3727/30/11/020} {\bibfield  {journal} {\bibinfo
   {journal} {J. Phys. D: Appl. Phys}\ }\textbf {\bibinfo {volume} {30}},\
  \bibinfo {pages} {1697} (\bibinfo {year} {1997})}\BibitemShut {NoStop}%
\bibitem [{\citenamefont {Herlach}(1999)}]{Herlach}%
  \BibitemOpen
  \bibfield  {author} {\bibinfo {author} {\bibfnamefont {F.}~\bibnamefont
  {Herlach}},\ }\href {https://doi.org/10.1088/0034-4885/62/6/201} {\bibfield
  {journal} {\bibinfo  {journal} {Rep. Prog. Phys.}\ }\textbf {\bibinfo
  {volume} {62}},\ \bibinfo {pages} {859} (\bibinfo {year} {1999})}\BibitemShut
  {NoStop}%
\bibitem [{\citenamefont {Miura}\ \emph {et~al.}(2003)\citenamefont {Miura},
  \citenamefont {Osada},\ and\ \citenamefont {Takeyama}}]{Miura}%
  \BibitemOpen
  \bibfield  {author} {\bibinfo {author} {\bibfnamefont {N.}~\bibnamefont
  {Miura}}, \bibinfo {author} {\bibfnamefont {T.}~\bibnamefont {Osada}},\ and\
  \bibinfo {author} {\bibfnamefont {S.}~\bibnamefont {Takeyama}},\ }\href
  {https://doi.org/10.1023/A:1025689218138} {\bibfield  {journal} {\bibinfo
  {journal} {J. Low Temp. Phys.}\ }\textbf {\bibinfo {volume} {133}},\ \bibinfo
  {pages} {139} (\bibinfo {year} {2003})}\BibitemShut {NoStop}%
\bibitem [{\citenamefont {Nakamura}\ \emph {et~al.}(2018)\citenamefont
  {Nakamura}, \citenamefont {Ikeda}, \citenamefont {Sawabe}, \citenamefont
  {Matsuda},\ and\ \citenamefont {Takeyama}}]{DN1200}%
  \BibitemOpen
  \bibfield  {author} {\bibinfo {author} {\bibfnamefont {D.}~\bibnamefont
  {Nakamura}}, \bibinfo {author} {\bibfnamefont {A.}~\bibnamefont {Ikeda}},
  \bibinfo {author} {\bibfnamefont {H.}~\bibnamefont {Sawabe}}, \bibinfo
  {author} {\bibfnamefont {Y.~H.}\ \bibnamefont {Matsuda}},\ and\ \bibinfo
  {author} {\bibfnamefont {S.}~\bibnamefont {Takeyama}},\ }\href
  {https://doi.org/10.1063/1.5044557} {\bibfield  {journal} {\bibinfo
  {journal} {Rev. Sci. Instrum.}\ }\textbf {\bibinfo {volume} {89}},\ \bibinfo
  {pages} {095106} (\bibinfo {year} {2018})}\BibitemShut {NoStop}%
\bibitem [{not()}]{note1}%
  \BibitemOpen
  \href@noop {} {}\bibinfo {note} {Note an exceptional technique where single
  shot XRD is utilized with a bright and broadband SR from a storage ring for
  shock physics experiments \cite{TurneaurePRL1, TurneaureSA, TurneaurePRL2,
  TurneaurePRL3, TurneaurePRL4, IchiyanagiSR}.}\BibitemShut {Stop}%
\bibitem [{\citenamefont {Ikeda}\ \emph {et~al.}({\natexlab{a}})\citenamefont
  {Ikeda}, \citenamefont {Matsuda}, \citenamefont {Nakamura}, \citenamefont
  {Takeyama}, \citenamefont {Tsuda}, \citenamefont {Nomura}, \citenamefont
  {Shimizu}, \citenamefont {Matsuo}, \citenamefont {Nomura}, \citenamefont
  {Kobayashi}, \citenamefont {Yajima}, \citenamefont {Ishikawa}, \citenamefont
  {Hiroi}, \citenamefont {Isobe}, \citenamefont {Yamauchi},\ and\ \citenamefont
  {Sato}}]{IkedaMG}%
  \BibitemOpen
  \bibfield  {author} {\bibinfo {author} {\bibfnamefont {A.}~\bibnamefont
  {Ikeda}}, \bibinfo {author} {\bibfnamefont {Y.~H.}\ \bibnamefont {Matsuda}},
  \bibinfo {author} {\bibfnamefont {D.}~\bibnamefont {Nakamura}}, \bibinfo
  {author} {\bibfnamefont {S.}~\bibnamefont {Takeyama}}, \bibinfo {author}
  {\bibfnamefont {H.}~\bibnamefont {Tsuda}}, \bibinfo {author} {\bibfnamefont
  {K.}~\bibnamefont {Nomura}}, \bibinfo {author} {\bibfnamefont
  {A.}~\bibnamefont {Shimizu}}, \bibinfo {author} {\bibfnamefont
  {A.}~\bibnamefont {Matsuo}}, \bibinfo {author} {\bibfnamefont
  {T.}~\bibnamefont {Nomura}}, \bibinfo {author} {\bibfnamefont {T.~C.}\
  \bibnamefont {Kobayashi}}, \bibinfo {author} {\bibfnamefont {T.}~\bibnamefont
  {Yajima}}, \bibinfo {author} {\bibfnamefont {H.}~\bibnamefont {Ishikawa}},
  \bibinfo {author} {\bibfnamefont {Z.}~\bibnamefont {Hiroi}}, \bibinfo
  {author} {\bibfnamefont {M.}~\bibnamefont {Isobe}}, \bibinfo {author}
  {\bibfnamefont {T.}~\bibnamefont {Yamauchi}},\ and\ \bibinfo {author}
  {\bibfnamefont {K.}~\bibnamefont {Sato}},\ }in\ \href
  {https://doi.org/10.1109/MEGAGAUSS.2018.8722656} {\emph {\bibinfo {booktitle}
  {2018 16th International Conference on Megagauss Magnetic Field Generation
  and Related Topics (MEGAGAUSS)}}},\ pp.\ \bibinfo {pages} {1--9}\BibitemShut
  {NoStop}%
\bibitem [{\citenamefont {Ikeda}\ \emph {et~al.}(2017)\citenamefont {Ikeda},
  \citenamefont {Nomura}, \citenamefont {Matsuda}, \citenamefont {Tani},
  \citenamefont {Kobayashi}, \citenamefont {Watanabe},\ and\ \citenamefont
  {Sato}}]{IkedaFBG2017}%
  \BibitemOpen
  \bibfield  {author} {\bibinfo {author} {\bibfnamefont {A.}~\bibnamefont
  {Ikeda}}, \bibinfo {author} {\bibfnamefont {T.}~\bibnamefont {Nomura}},
  \bibinfo {author} {\bibfnamefont {Y.~H.}\ \bibnamefont {Matsuda}}, \bibinfo
  {author} {\bibfnamefont {S.}~\bibnamefont {Tani}}, \bibinfo {author}
  {\bibfnamefont {Y.}~\bibnamefont {Kobayashi}}, \bibinfo {author}
  {\bibfnamefont {H.}~\bibnamefont {Watanabe}},\ and\ \bibinfo {author}
  {\bibfnamefont {K.}~\bibnamefont {Sato}},\ }\href
  {https://doi.org/10.1063/1.4999452} {\bibfield  {journal} {\bibinfo
  {journal} {Rev. Sci. Instrum.}\ }\textbf {\bibinfo {volume} {88}},\ \bibinfo
  {pages} {083906} (\bibinfo {year} {2017})}\BibitemShut {NoStop}%
\bibitem [{\citenamefont {Nomura}\ \emph {et~al.}(2014)\citenamefont {Nomura},
  \citenamefont {Matsuda}, \citenamefont {Takeyama}, \citenamefont {Matsuo},
  \citenamefont {Kindo}, \citenamefont {Her},\ and\ \citenamefont
  {Kobayashi}}]{NomuraPRL2014}%
  \BibitemOpen
  \bibfield  {author} {\bibinfo {author} {\bibfnamefont {T.}~\bibnamefont
  {Nomura}}, \bibinfo {author} {\bibfnamefont {Y.~H.}\ \bibnamefont {Matsuda}},
  \bibinfo {author} {\bibfnamefont {S.}~\bibnamefont {Takeyama}}, \bibinfo
  {author} {\bibfnamefont {A.}~\bibnamefont {Matsuo}}, \bibinfo {author}
  {\bibfnamefont {K.}~\bibnamefont {Kindo}}, \bibinfo {author} {\bibfnamefont
  {J.~L.}\ \bibnamefont {Her}},\ and\ \bibinfo {author} {\bibfnamefont {T.~C.}\
  \bibnamefont {Kobayashi}},\ }\href
  {https://doi.org/10.1103/PhysRevLett.112.247201} {\bibfield  {journal}
  {\bibinfo  {journal} {Phys. Rev. Lett.}\ }\textbf {\bibinfo {volume} {112}},\
  \bibinfo {pages} {247201} (\bibinfo {year} {2014})}\BibitemShut {NoStop}%
\bibitem [{\citenamefont {Ikeda}\ \emph {et~al.}(2016)\citenamefont {Ikeda},
  \citenamefont {Nomura}, \citenamefont {Matsuda}, \citenamefont {Matsuo},
  \citenamefont {Kindo},\ and\ \citenamefont {Sato}}]{IkedaLCO}%
  \BibitemOpen
  \bibfield  {author} {\bibinfo {author} {\bibfnamefont {A.}~\bibnamefont
  {Ikeda}}, \bibinfo {author} {\bibfnamefont {T.}~\bibnamefont {Nomura}},
  \bibinfo {author} {\bibfnamefont {Y.~H.}\ \bibnamefont {Matsuda}}, \bibinfo
  {author} {\bibfnamefont {A.}~\bibnamefont {Matsuo}}, \bibinfo {author}
  {\bibfnamefont {K.}~\bibnamefont {Kindo}},\ and\ \bibinfo {author}
  {\bibfnamefont {K.}~\bibnamefont {Sato}},\ }\href
  {https://doi.org/10.1103/PhysRevB.93.220401} {\bibfield  {journal} {\bibinfo
  {journal} {Phys. Rev. B}\ }\textbf {\bibinfo {volume} {93}},\ \bibinfo
  {pages} {220401(R)} (\bibinfo {year} {2016})}\BibitemShut {NoStop}%
\bibitem [{\citenamefont {Ikeda}\ \emph {et~al.}({\natexlab{b}})\citenamefont
  {Ikeda}, \citenamefont {Matsuda},\ and\ \citenamefont {Sato}}]{IkedaArxiv}%
  \BibitemOpen
  \bibfield  {author} {\bibinfo {author} {\bibfnamefont {A.}~\bibnamefont
  {Ikeda}}, \bibinfo {author} {\bibfnamefont {Y.~H.}\ \bibnamefont {Matsuda}},\
  and\ \bibinfo {author} {\bibfnamefont {K.}~\bibnamefont {Sato}},\ }\href@noop
  {} {} ({\natexlab{b}}),\ \Eprint {https://arxiv.org/abs/arXiv:2004.12395}
  {arXiv:2004.12395} \BibitemShut {NoStop}%
\bibitem [{\citenamefont {Miyata}\ \emph {et~al.}(2011)\citenamefont {Miyata},
  \citenamefont {Ueda}, \citenamefont {Ueda}, \citenamefont {Sawabe},\ and\
  \citenamefont {Takeyama}}]{MiyataPRL}%
  \BibitemOpen
  \bibfield  {author} {\bibinfo {author} {\bibfnamefont {A.}~\bibnamefont
  {Miyata}}, \bibinfo {author} {\bibfnamefont {H.}~\bibnamefont {Ueda}},
  \bibinfo {author} {\bibfnamefont {Y.}~\bibnamefont {Ueda}}, \bibinfo {author}
  {\bibfnamefont {H.}~\bibnamefont {Sawabe}},\ and\ \bibinfo {author}
  {\bibfnamefont {S.}~\bibnamefont {Takeyama}},\ }\href
  {https://doi.org/10.1103/PhysRevLett.107.207203} {\bibfield  {journal}
  {\bibinfo  {journal} {Phys. Rev. Lett.}\ }\textbf {\bibinfo {volume} {107}},\
  \bibinfo {pages} {207203} (\bibinfo {year} {2011})}\BibitemShut {NoStop}%
\bibitem [{\citenamefont {Terashima}\ \emph {et~al.}(2017)\citenamefont
  {Terashima}, \citenamefont {Ikeda}, \citenamefont {Matsuda}, \citenamefont
  {Kondo}, \citenamefont {Kindo},\ and\ \citenamefont {Iga}}]{TerashimaJPSJ}%
  \BibitemOpen
  \bibfield  {author} {\bibinfo {author} {\bibfnamefont {T.~T.}\ \bibnamefont
  {Terashima}}, \bibinfo {author} {\bibfnamefont {A.}~\bibnamefont {Ikeda}},
  \bibinfo {author} {\bibfnamefont {Y.~H.}\ \bibnamefont {Matsuda}}, \bibinfo
  {author} {\bibfnamefont {A.}~\bibnamefont {Kondo}}, \bibinfo {author}
  {\bibfnamefont {K.}~\bibnamefont {Kindo}},\ and\ \bibinfo {author}
  {\bibfnamefont {F.}~\bibnamefont {Iga}},\ }\href
  {https://doi.org/10.7566/JPSJ.86.054710} {\bibfield  {journal} {\bibinfo
  {journal} {J. Phys. Soc. Jpn.}\ }\textbf {\bibinfo {volume} {86}},\ \bibinfo
  {pages} {054710} (\bibinfo {year} {2017})}\BibitemShut {NoStop}%
\bibitem [{\citenamefont {Matsuda}\ \emph {et~al.}(2020)\citenamefont
  {Matsuda}, \citenamefont {Nakamura}, \citenamefont {Ikeda}, \citenamefont
  {Takeyama}, \citenamefont {Suga}, \citenamefont {Nakahara},\ and\
  \citenamefont {Muraoka}}]{YMatsudaNC}%
  \BibitemOpen
  \bibfield  {author} {\bibinfo {author} {\bibfnamefont {Y.~H.}\ \bibnamefont
  {Matsuda}}, \bibinfo {author} {\bibfnamefont {D.}~\bibnamefont {Nakamura}},
  \bibinfo {author} {\bibfnamefont {A.}~\bibnamefont {Ikeda}}, \bibinfo
  {author} {\bibfnamefont {S.}~\bibnamefont {Takeyama}}, \bibinfo {author}
  {\bibfnamefont {Y.}~\bibnamefont {Suga}}, \bibinfo {author} {\bibfnamefont
  {H.}~\bibnamefont {Nakahara}},\ and\ \bibinfo {author} {\bibfnamefont
  {Y.}~\bibnamefont {Muraoka}},\ }\href
  {https://doi.org/10.1038/s41467-020-17416-w} {\bibfield  {journal} {\bibinfo
  {journal} {Nat. Commun.}\ }\textbf {\bibinfo {volume} {11}},\ \bibinfo
  {pages} {3591} (\bibinfo {year} {2020})}\BibitemShut {NoStop}%
\bibitem [{\citenamefont {Ishikawa}\ \emph {et~al.}(2012)\citenamefont
  {Ishikawa}, \citenamefont {Aoyagi}, \citenamefont {Asaka}, \citenamefont
  {Asano}, \citenamefont {Azumi}, \citenamefont {Bizen}, \citenamefont {Ego},
  \citenamefont {Fukami}, \citenamefont {Fukui}, \citenamefont {Furukawa},
  \citenamefont {Goto}, \citenamefont {Hanaki}, \citenamefont {Hara},
  \citenamefont {Hasegawa}, \citenamefont {Hatsui}, \citenamefont {Higashiya},
  \citenamefont {Hirono}, \citenamefont {Hosoda}, \citenamefont {Ishii},
  \citenamefont {Inagaki}, \citenamefont {Inubushi}, \citenamefont {Itoga},
  \citenamefont {Joti}, \citenamefont {Kago}, \citenamefont {Kameshima},
  \citenamefont {Kimura}, \citenamefont {Kirihara}, \citenamefont {Kiyomichi},
  \citenamefont {Kobayashi}, \citenamefont {Kondo}, \citenamefont {Kudo},
  \citenamefont {Maesaka}, \citenamefont {Maréchal}, \citenamefont {Masuda},
  \citenamefont {Matsubara}, \citenamefont {Matsumoto}, \citenamefont
  {Matsushita}, \citenamefont {Matsui}, \citenamefont {Nagasono}, \citenamefont
  {Nariyama}, \citenamefont {Ohashi}, \citenamefont {Ohata}, \citenamefont
  {Ohshima}, \citenamefont {Ono}, \citenamefont {Otake}, \citenamefont {Saji},
  \citenamefont {Sakurai}, \citenamefont {Sato}, \citenamefont {Sawada},
  \citenamefont {Seike}, \citenamefont {Shirasawa}, \citenamefont {Sugimoto},
  \citenamefont {Suzuki}, \citenamefont {Takahashi}, \citenamefont {Takebe},
  \citenamefont {Takeshita}, \citenamefont {Tamasaku}, \citenamefont {Tanaka},
  \citenamefont {Tanaka}, \citenamefont {Tanaka}, \citenamefont {Togashi},
  \citenamefont {Togawa}, \citenamefont {Tokuhisa}, \citenamefont {Tomizawa},
  \citenamefont {Tono}, \citenamefont {Wu}, \citenamefont {Yabashi},
  \citenamefont {Yamaga}, \citenamefont {Yamashita}, \citenamefont {Yanagida},
  \citenamefont {Zhang}, \citenamefont {Shintake}, \citenamefont {Kitamura},\
  and\ \citenamefont {Kumagai}}]{IshikawaNP}%
  \BibitemOpen
  \bibfield  {author} {\bibinfo {author} {\bibfnamefont {T.}~\bibnamefont
  {Ishikawa}}, \bibinfo {author} {\bibfnamefont {H.}~\bibnamefont {Aoyagi}},
  \bibinfo {author} {\bibfnamefont {T.}~\bibnamefont {Asaka}}, \bibinfo
  {author} {\bibfnamefont {Y.}~\bibnamefont {Asano}}, \bibinfo {author}
  {\bibfnamefont {N.}~\bibnamefont {Azumi}}, \bibinfo {author} {\bibfnamefont
  {T.}~\bibnamefont {Bizen}}, \bibinfo {author} {\bibfnamefont
  {H.}~\bibnamefont {Ego}}, \bibinfo {author} {\bibfnamefont {K.}~\bibnamefont
  {Fukami}}, \bibinfo {author} {\bibfnamefont {T.}~\bibnamefont {Fukui}},
  \bibinfo {author} {\bibfnamefont {Y.}~\bibnamefont {Furukawa}}, \bibinfo
  {author} {\bibfnamefont {S.}~\bibnamefont {Goto}}, \bibinfo {author}
  {\bibfnamefont {H.}~\bibnamefont {Hanaki}}, \bibinfo {author} {\bibfnamefont
  {T.}~\bibnamefont {Hara}}, \bibinfo {author} {\bibfnamefont {T.}~\bibnamefont
  {Hasegawa}}, \bibinfo {author} {\bibfnamefont {T.}~\bibnamefont {Hatsui}},
  \bibinfo {author} {\bibfnamefont {A.}~\bibnamefont {Higashiya}}, \bibinfo
  {author} {\bibfnamefont {T.}~\bibnamefont {Hirono}}, \bibinfo {author}
  {\bibfnamefont {N.}~\bibnamefont {Hosoda}}, \bibinfo {author} {\bibfnamefont
  {M.}~\bibnamefont {Ishii}}, \bibinfo {author} {\bibfnamefont
  {T.}~\bibnamefont {Inagaki}}, \bibinfo {author} {\bibfnamefont
  {Y.}~\bibnamefont {Inubushi}}, \bibinfo {author} {\bibfnamefont
  {T.}~\bibnamefont {Itoga}}, \bibinfo {author} {\bibfnamefont
  {Y.}~\bibnamefont {Joti}}, \bibinfo {author} {\bibfnamefont {M.}~\bibnamefont
  {Kago}}, \bibinfo {author} {\bibfnamefont {T.}~\bibnamefont {Kameshima}},
  \bibinfo {author} {\bibfnamefont {H.}~\bibnamefont {Kimura}}, \bibinfo
  {author} {\bibfnamefont {Y.}~\bibnamefont {Kirihara}}, \bibinfo {author}
  {\bibfnamefont {A.}~\bibnamefont {Kiyomichi}}, \bibinfo {author}
  {\bibfnamefont {T.}~\bibnamefont {Kobayashi}}, \bibinfo {author}
  {\bibfnamefont {C.}~\bibnamefont {Kondo}}, \bibinfo {author} {\bibfnamefont
  {T.}~\bibnamefont {Kudo}}, \bibinfo {author} {\bibfnamefont {H.}~\bibnamefont
  {Maesaka}}, \bibinfo {author} {\bibfnamefont {X.~M.}\ \bibnamefont
  {Maréchal}}, \bibinfo {author} {\bibfnamefont {T.}~\bibnamefont {Masuda}},
  \bibinfo {author} {\bibfnamefont {S.}~\bibnamefont {Matsubara}}, \bibinfo
  {author} {\bibfnamefont {T.}~\bibnamefont {Matsumoto}}, \bibinfo {author}
  {\bibfnamefont {T.}~\bibnamefont {Matsushita}}, \bibinfo {author}
  {\bibfnamefont {S.}~\bibnamefont {Matsui}}, \bibinfo {author} {\bibfnamefont
  {M.}~\bibnamefont {Nagasono}}, \bibinfo {author} {\bibfnamefont
  {N.}~\bibnamefont {Nariyama}}, \bibinfo {author} {\bibfnamefont
  {H.}~\bibnamefont {Ohashi}}, \bibinfo {author} {\bibfnamefont
  {T.}~\bibnamefont {Ohata}}, \bibinfo {author} {\bibfnamefont
  {T.}~\bibnamefont {Ohshima}}, \bibinfo {author} {\bibfnamefont
  {S.}~\bibnamefont {Ono}}, \bibinfo {author} {\bibfnamefont {Y.}~\bibnamefont
  {Otake}}, \bibinfo {author} {\bibfnamefont {C.}~\bibnamefont {Saji}},
  \bibinfo {author} {\bibfnamefont {T.}~\bibnamefont {Sakurai}}, \bibinfo
  {author} {\bibfnamefont {T.}~\bibnamefont {Sato}}, \bibinfo {author}
  {\bibfnamefont {K.}~\bibnamefont {Sawada}}, \bibinfo {author} {\bibfnamefont
  {T.}~\bibnamefont {Seike}}, \bibinfo {author} {\bibfnamefont
  {K.}~\bibnamefont {Shirasawa}}, \bibinfo {author} {\bibfnamefont
  {T.}~\bibnamefont {Sugimoto}}, \bibinfo {author} {\bibfnamefont
  {S.}~\bibnamefont {Suzuki}}, \bibinfo {author} {\bibfnamefont
  {S.}~\bibnamefont {Takahashi}}, \bibinfo {author} {\bibfnamefont
  {H.}~\bibnamefont {Takebe}}, \bibinfo {author} {\bibfnamefont
  {K.}~\bibnamefont {Takeshita}}, \bibinfo {author} {\bibfnamefont
  {K.}~\bibnamefont {Tamasaku}}, \bibinfo {author} {\bibfnamefont
  {H.}~\bibnamefont {Tanaka}}, \bibinfo {author} {\bibfnamefont
  {R.}~\bibnamefont {Tanaka}}, \bibinfo {author} {\bibfnamefont
  {T.}~\bibnamefont {Tanaka}}, \bibinfo {author} {\bibfnamefont
  {T.}~\bibnamefont {Togashi}}, \bibinfo {author} {\bibfnamefont
  {K.}~\bibnamefont {Togawa}}, \bibinfo {author} {\bibfnamefont
  {A.}~\bibnamefont {Tokuhisa}}, \bibinfo {author} {\bibfnamefont
  {H.}~\bibnamefont {Tomizawa}}, \bibinfo {author} {\bibfnamefont
  {K.}~\bibnamefont {Tono}}, \bibinfo {author} {\bibfnamefont {S.}~\bibnamefont
  {Wu}}, \bibinfo {author} {\bibfnamefont {M.}~\bibnamefont {Yabashi}},
  \bibinfo {author} {\bibfnamefont {M.}~\bibnamefont {Yamaga}}, \bibinfo
  {author} {\bibfnamefont {A.}~\bibnamefont {Yamashita}}, \bibinfo {author}
  {\bibfnamefont {K.}~\bibnamefont {Yanagida}}, \bibinfo {author}
  {\bibfnamefont {C.}~\bibnamefont {Zhang}}, \bibinfo {author} {\bibfnamefont
  {T.}~\bibnamefont {Shintake}}, \bibinfo {author} {\bibfnamefont
  {H.}~\bibnamefont {Kitamura}},\ and\ \bibinfo {author} {\bibfnamefont
  {N.}~\bibnamefont {Kumagai}},\ }\href
  {https://doi.org/10.1038/nphoton.2012.141} {\bibfield  {journal} {\bibinfo
  {journal} {Nat. Photonics}\ }\textbf {\bibinfo {volume} {6}},\ \bibinfo
  {pages} {540} (\bibinfo {year} {2012})}\BibitemShut {NoStop}%
\bibitem [{\citenamefont {Yabashi}\ \emph {et~al.}(2015)\citenamefont
  {Yabashi}, \citenamefont {Tanaka},\ and\ \citenamefont
  {Ishikawa}}]{YabashiJSR}%
  \BibitemOpen
  \bibfield  {author} {\bibinfo {author} {\bibfnamefont {M.}~\bibnamefont
  {Yabashi}}, \bibinfo {author} {\bibfnamefont {H.}~\bibnamefont {Tanaka}},\
  and\ \bibinfo {author} {\bibfnamefont {T.}~\bibnamefont {Ishikawa}},\ }\href
  {https://doi.org/10.1107/S1600577515004658} {\bibfield  {journal} {\bibinfo
  {journal} {J. Synchrotron Rad.}\ }\textbf {\bibinfo {volume} {22}},\ \bibinfo
  {pages} {477} (\bibinfo {year} {2015})}\BibitemShut {NoStop}%
\bibitem [{\citenamefont {Kameshima}\ \emph {et~al.}(2014)\citenamefont
  {Kameshima}, \citenamefont {Ono}, \citenamefont {Kudo}, \citenamefont
  {Ozaki}, \citenamefont {Kirihara}, \citenamefont {Kobayashi}, \citenamefont
  {Inubushi}, \citenamefont {Yabashi}, \citenamefont {Horigome}, \citenamefont
  {Holland}, \citenamefont {Holland}, \citenamefont {Burt}, \citenamefont
  {Murao},\ and\ \citenamefont {Hatsui}}]{Kameshima}%
  \BibitemOpen
  \bibfield  {author} {\bibinfo {author} {\bibfnamefont {T.}~\bibnamefont
  {Kameshima}}, \bibinfo {author} {\bibfnamefont {S.}~\bibnamefont {Ono}},
  \bibinfo {author} {\bibfnamefont {T.}~\bibnamefont {Kudo}}, \bibinfo {author}
  {\bibfnamefont {K.}~\bibnamefont {Ozaki}}, \bibinfo {author} {\bibfnamefont
  {Y.}~\bibnamefont {Kirihara}}, \bibinfo {author} {\bibfnamefont
  {K.}~\bibnamefont {Kobayashi}}, \bibinfo {author} {\bibfnamefont
  {Y.}~\bibnamefont {Inubushi}}, \bibinfo {author} {\bibfnamefont
  {M.}~\bibnamefont {Yabashi}}, \bibinfo {author} {\bibfnamefont
  {T.}~\bibnamefont {Horigome}}, \bibinfo {author} {\bibfnamefont
  {A.}~\bibnamefont {Holland}}, \bibinfo {author} {\bibfnamefont
  {K.}~\bibnamefont {Holland}}, \bibinfo {author} {\bibfnamefont
  {D.}~\bibnamefont {Burt}}, \bibinfo {author} {\bibfnamefont {H.}~\bibnamefont
  {Murao}},\ and\ \bibinfo {author} {\bibfnamefont {T.}~\bibnamefont
  {Hatsui}},\ }\href {https://doi.org/10.1063/1.4867668} {\bibfield  {journal}
  {\bibinfo  {journal} {Rev. Sci. Instrum.}\ }\textbf {\bibinfo {volume}
  {85}},\ \bibinfo {pages} {033110} (\bibinfo {year} {2014})}\BibitemShut
  {NoStop}%
\bibitem [{\citenamefont {H.~Matsuda}\ \emph {et~al.}(2007)\citenamefont
  {H.~Matsuda}, \citenamefont {Inami}, \citenamefont {Ohwada}, \citenamefont
  {Murata}, \citenamefont {Nojiri}, \citenamefont {Murakami}, \citenamefont
  {Ohta}, \citenamefont {Zhang},\ and\ \citenamefont {Yoshimura}}]{YHM2007}%
  \BibitemOpen
  \bibfield  {author} {\bibinfo {author} {\bibfnamefont {Y.}~\bibnamefont
  {H.~Matsuda}}, \bibinfo {author} {\bibfnamefont {T.}~\bibnamefont {Inami}},
  \bibinfo {author} {\bibfnamefont {K.}~\bibnamefont {Ohwada}}, \bibinfo
  {author} {\bibfnamefont {Y.}~\bibnamefont {Murata}}, \bibinfo {author}
  {\bibfnamefont {H.}~\bibnamefont {Nojiri}}, \bibinfo {author} {\bibfnamefont
  {Y.}~\bibnamefont {Murakami}}, \bibinfo {author} {\bibfnamefont
  {H.}~\bibnamefont {Ohta}}, \bibinfo {author} {\bibfnamefont {W.}~\bibnamefont
  {Zhang}},\ and\ \bibinfo {author} {\bibfnamefont {K.}~\bibnamefont
  {Yoshimura}},\ }\href {https://doi.org/10.1143/jpsj.76.034702} {\bibfield
  {journal} {\bibinfo  {journal} {J. Phys. Soc. Jpn.}\ }\textbf {\bibinfo
  {volume} {76}},\ \bibinfo {pages} {034702} (\bibinfo {year}
  {2007})}\BibitemShut {NoStop}%
\bibitem [{\citenamefont {Tokura}\ and\ \citenamefont
  {Tomioka}(1999)}]{Tokura}%
  \BibitemOpen
  \bibfield  {author} {\bibinfo {author} {\bibfnamefont {Y.}~\bibnamefont
  {Tokura}}\ and\ \bibinfo {author} {\bibfnamefont {Y.}~\bibnamefont
  {Tomioka}},\ }\href
  {https://doi.org/http://dx.doi.org/10.1016/S0304-8853(99)00352-2} {\bibfield
  {journal} {\bibinfo  {journal} {J. Mag. Mag. Mat.}\ }\textbf {\bibinfo
  {volume} {200}},\ \bibinfo {pages} {1} (\bibinfo {year} {1999})}\BibitemShut
  {NoStop}%
\bibitem [{\citenamefont {Tomioka}\ \emph {et~al.}(1996)\citenamefont
  {Tomioka}, \citenamefont {Asamitsu}, \citenamefont {Kuwahara}, \citenamefont
  {Moritomo},\ and\ \citenamefont {Tokura}}]{Tomioka}%
  \BibitemOpen
  \bibfield  {author} {\bibinfo {author} {\bibfnamefont {Y.}~\bibnamefont
  {Tomioka}}, \bibinfo {author} {\bibfnamefont {A.}~\bibnamefont {Asamitsu}},
  \bibinfo {author} {\bibfnamefont {H.}~\bibnamefont {Kuwahara}}, \bibinfo
  {author} {\bibfnamefont {Y.}~\bibnamefont {Moritomo}},\ and\ \bibinfo
  {author} {\bibfnamefont {Y.}~\bibnamefont {Tokura}},\ }\href
  {https://doi.org/10.1103/PhysRevB.53.R1689} {\bibfield  {journal} {\bibinfo
  {journal} {Phys. Rev. B}\ }\textbf {\bibinfo {volume} {53}},\ \bibinfo
  {pages} {R1689} (\bibinfo {year} {1996})}\BibitemShut {NoStop}%
\bibitem [{\citenamefont {Kuwahara}\ \emph {et~al.}(1995)\citenamefont
  {Kuwahara}, \citenamefont {Tomioka}, \citenamefont {Asamitsu}, \citenamefont
  {Moritomo},\ and\ \citenamefont {Tokura}}]{Kuwahara}%
  \BibitemOpen
  \bibfield  {author} {\bibinfo {author} {\bibfnamefont {H.}~\bibnamefont
  {Kuwahara}}, \bibinfo {author} {\bibfnamefont {Y.}~\bibnamefont {Tomioka}},
  \bibinfo {author} {\bibfnamefont {A.}~\bibnamefont {Asamitsu}}, \bibinfo
  {author} {\bibfnamefont {Y.}~\bibnamefont {Moritomo}},\ and\ \bibinfo
  {author} {\bibfnamefont {Y.}~\bibnamefont {Tokura}},\ }\href
  {https://doi.org/10.1126/science.270.5238.961} {\bibfield  {journal}
  {\bibinfo  {journal} {Science}\ }\textbf {\bibinfo {volume} {270}},\ \bibinfo
  {pages} {961} (\bibinfo {year} {1995})}\BibitemShut {NoStop}%
\bibitem [{\citenamefont {Inoue}\ \emph {et~al.}(2019)\citenamefont {Inoue},
  \citenamefont {Osaka}, \citenamefont {Hara}, \citenamefont {Tanaka},
  \citenamefont {Inagaki}, \citenamefont {Fukui}, \citenamefont {Goto},
  \citenamefont {Inubushi}, \citenamefont {Kimura}, \citenamefont {Kinjo},
  \citenamefont {Ohashi}, \citenamefont {Togawa}, \citenamefont {Tono},
  \citenamefont {Yamaga}, \citenamefont {Tanaka}, \citenamefont {Ishikawa},\
  and\ \citenamefont {Yabashi}}]{InoueNP}%
  \BibitemOpen
  \bibfield  {author} {\bibinfo {author} {\bibfnamefont {I.}~\bibnamefont
  {Inoue}}, \bibinfo {author} {\bibfnamefont {T.}~\bibnamefont {Osaka}},
  \bibinfo {author} {\bibfnamefont {T.}~\bibnamefont {Hara}}, \bibinfo {author}
  {\bibfnamefont {T.}~\bibnamefont {Tanaka}}, \bibinfo {author} {\bibfnamefont
  {T.}~\bibnamefont {Inagaki}}, \bibinfo {author} {\bibfnamefont
  {T.}~\bibnamefont {Fukui}}, \bibinfo {author} {\bibfnamefont
  {S.}~\bibnamefont {Goto}}, \bibinfo {author} {\bibfnamefont {Y.}~\bibnamefont
  {Inubushi}}, \bibinfo {author} {\bibfnamefont {H.}~\bibnamefont {Kimura}},
  \bibinfo {author} {\bibfnamefont {R.}~\bibnamefont {Kinjo}}, \bibinfo
  {author} {\bibfnamefont {H.}~\bibnamefont {Ohashi}}, \bibinfo {author}
  {\bibfnamefont {K.}~\bibnamefont {Togawa}}, \bibinfo {author} {\bibfnamefont
  {K.}~\bibnamefont {Tono}}, \bibinfo {author} {\bibfnamefont {M.}~\bibnamefont
  {Yamaga}}, \bibinfo {author} {\bibfnamefont {H.}~\bibnamefont {Tanaka}},
  \bibinfo {author} {\bibfnamefont {T.}~\bibnamefont {Ishikawa}},\ and\
  \bibinfo {author} {\bibfnamefont {M.}~\bibnamefont {Yabashi}},\ }\href
  {https://doi.org/10.1038/s41566-019-0365-y} {\bibfield  {journal} {\bibinfo
  {journal} {Nat. Photonics}\ }\textbf {\bibinfo {volume} {13}},\ \bibinfo
  {pages} {319} (\bibinfo {year} {2019})}\BibitemShut {NoStop}%
\bibitem [{\citenamefont {Turneaure}\ \emph {et~al.}(2016)\citenamefont
  {Turneaure}, \citenamefont {Sinclair},\ and\ \citenamefont
  {Gupta}}]{TurneaurePRL1}%
  \BibitemOpen
  \bibfield  {author} {\bibinfo {author} {\bibfnamefont {S.~J.}\ \bibnamefont
  {Turneaure}}, \bibinfo {author} {\bibfnamefont {N.}~\bibnamefont
  {Sinclair}},\ and\ \bibinfo {author} {\bibfnamefont {Y.~M.}\ \bibnamefont
  {Gupta}},\ }\href {https://doi.org/10.1103/PhysRevLett.117.045502} {\bibfield
   {journal} {\bibinfo  {journal} {Phys. Rev. Lett.}\ }\textbf {\bibinfo
  {volume} {117}},\ \bibinfo {pages} {045502} (\bibinfo {year}
  {2016})}\BibitemShut {NoStop}%
\bibitem [{\citenamefont {Turneaure}\ \emph {et~al.}(2017)\citenamefont
  {Turneaure}, \citenamefont {Sharma}, \citenamefont {Volz}, \citenamefont
  {Winey},\ and\ \citenamefont {Gupta}}]{TurneaureSA}%
  \BibitemOpen
  \bibfield  {author} {\bibinfo {author} {\bibfnamefont {S.~J.}\ \bibnamefont
  {Turneaure}}, \bibinfo {author} {\bibfnamefont {S.~M.}\ \bibnamefont
  {Sharma}}, \bibinfo {author} {\bibfnamefont {T.~J.}\ \bibnamefont {Volz}},
  \bibinfo {author} {\bibfnamefont {J.~M.}\ \bibnamefont {Winey}},\ and\
  \bibinfo {author} {\bibfnamefont {Y.~M.}\ \bibnamefont {Gupta}},\ }\href
  {https://doi.org/10.1126/sciadv.aao3561} {\bibfield  {journal} {\bibinfo
  {journal} {Sci. Adv.}\ }\textbf {\bibinfo {volume} {3}},\ \bibinfo {pages}
  {eaao3561} (\bibinfo {year} {2017})}\BibitemShut {NoStop}%
\bibitem [{\citenamefont {Tracy}\ \emph {et~al.}(2018)\citenamefont {Tracy},
  \citenamefont {Turneaure},\ and\ \citenamefont {Duffy}}]{TurneaurePRL2}%
  \BibitemOpen
  \bibfield  {author} {\bibinfo {author} {\bibfnamefont {S.~J.}\ \bibnamefont
  {Tracy}}, \bibinfo {author} {\bibfnamefont {S.~J.}\ \bibnamefont
  {Turneaure}},\ and\ \bibinfo {author} {\bibfnamefont {T.~S.}\ \bibnamefont
  {Duffy}},\ }\bibfield  {title} {\bibinfo {title} {In situ x-ray diffraction
  of shock-compressed fused silica},\ }\href
  {https://doi.org/10.1103/PhysRevLett.120.135702} {\bibfield  {journal}
  {\bibinfo  {journal} {Phys. Rev. Lett.}\ }\textbf {\bibinfo {volume} {120}},\
  \bibinfo {pages} {135702} (\bibinfo {year} {2018})}\BibitemShut {NoStop}%
\bibitem [{\citenamefont {Turneaure}\ \emph
  {et~al.}(2018{\natexlab{a}})\citenamefont {Turneaure}, \citenamefont
  {Renganathan}, \citenamefont {Winey},\ and\ \citenamefont
  {Gupta}}]{TurneaurePRL3}%
  \BibitemOpen
  \bibfield  {author} {\bibinfo {author} {\bibfnamefont {S.~J.}\ \bibnamefont
  {Turneaure}}, \bibinfo {author} {\bibfnamefont {P.}~\bibnamefont
  {Renganathan}}, \bibinfo {author} {\bibfnamefont {J.~M.}\ \bibnamefont
  {Winey}},\ and\ \bibinfo {author} {\bibfnamefont {Y.~M.}\ \bibnamefont
  {Gupta}},\ }\href {https://doi.org/10.1103/PhysRevLett.120.265503} {\bibfield
   {journal} {\bibinfo  {journal} {Phys. Rev. Lett.}\ }\textbf {\bibinfo
  {volume} {120}},\ \bibinfo {pages} {265503} (\bibinfo {year}
  {2018}{\natexlab{a}})}\BibitemShut {NoStop}%
\bibitem [{\citenamefont {Turneaure}\ \emph
  {et~al.}(2018{\natexlab{b}})\citenamefont {Turneaure}, \citenamefont
  {Sharma},\ and\ \citenamefont {Gupta}}]{TurneaurePRL4}%
  \BibitemOpen
  \bibfield  {author} {\bibinfo {author} {\bibfnamefont {S.~J.}\ \bibnamefont
  {Turneaure}}, \bibinfo {author} {\bibfnamefont {S.~M.}\ \bibnamefont
  {Sharma}},\ and\ \bibinfo {author} {\bibfnamefont {Y.~M.}\ \bibnamefont
  {Gupta}},\ }\href {https://doi.org/10.1103/PhysRevLett.121.135701} {\bibfield
   {journal} {\bibinfo  {journal} {Phys. Rev. Lett.}\ }\textbf {\bibinfo
  {volume} {121}},\ \bibinfo {pages} {135701} (\bibinfo {year}
  {2018}{\natexlab{b}})}\BibitemShut {NoStop}%
\bibitem [{\citenamefont {Ichiyanagi}\ \emph {et~al.}(2019)\citenamefont
  {Ichiyanagi}, \citenamefont {Takagi}, \citenamefont {Kawai}, \citenamefont
  {Fukaya}, \citenamefont {Nozawa}, \citenamefont {Nakamura}, \citenamefont
  {Liss}, \citenamefont {Kimura},\ and\ \citenamefont {Adachi}}]{IchiyanagiSR}%
  \BibitemOpen
  \bibfield  {author} {\bibinfo {author} {\bibfnamefont {K.}~\bibnamefont
  {Ichiyanagi}}, \bibinfo {author} {\bibfnamefont {S.}~\bibnamefont {Takagi}},
  \bibinfo {author} {\bibfnamefont {N.}~\bibnamefont {Kawai}}, \bibinfo
  {author} {\bibfnamefont {R.}~\bibnamefont {Fukaya}}, \bibinfo {author}
  {\bibfnamefont {S.}~\bibnamefont {Nozawa}}, \bibinfo {author} {\bibfnamefont
  {K.~G.}\ \bibnamefont {Nakamura}}, \bibinfo {author} {\bibfnamefont {K.~D.}\
  \bibnamefont {Liss}}, \bibinfo {author} {\bibfnamefont {M.}~\bibnamefont
  {Kimura}},\ and\ \bibinfo {author} {\bibfnamefont {S.~i.}\ \bibnamefont
  {Adachi}},\ }\href {https://doi.org/10.1038/s41598-019-43876-2} {\bibfield
  {journal} {\bibinfo  {journal} {Sci. Rep.}\ }\textbf {\bibinfo {volume}
  {9}},\ \bibinfo {pages} {7604} (\bibinfo {year} {2019})}\BibitemShut
  {NoStop}%
\end{thebibliography}%
\end{document}